\documentclass[aps,floatfix,onecolumn,a4paper,noshowpacs, nofootinbib,superscriptaddress,11pt]{revtex4}
\usepackage{caption}
\usepackage{caption}
\usepackage{booktabs}
\usepackage{braket}

\usepackage{multirow}
\usepackage{multirow,array}
\usepackage[utf8]{inputenc}
\usepackage{amsmath,amssymb,graphicx,bm}
\usepackage{mathrsfs}
\usepackage{geometry}
\usepackage{caption}
\newcommand{\dd}{\mathrm{d}}

\usepackage{siunitx}
\usepackage{hyperref}
\hypersetup{colorlinks=true, linkcolor=blue, citecolor=blue, urlcolor=blue}
\usepackage{graphicx,float,tikz}
\usepackage[all]{xy}
\usepackage{bm,amsmath,upgreek,bigints}

\newcommand{\lp}{\ell_{\rm p}}
\newcommand{\mpl}{m_{\rm p}}
\newcommand{\gn}{G_{\rm N}}

\usepackage{amssymb}
\usepackage{color}
\usepackage{epsfig,bm}		
\usepackage{graphicx,epstopdf}
\usepackage{subfigure,upgreek}
\usepackage{pdfpages}
\usepackage{multirow}

\newcommand{\beq}{\begin{eqnarray}}
\newcommand{\eeq}{\end{eqnarray}}

\usepackage{bookmark,textgreek}
\usepackage{hyperref,color,xcolor}
\definecolor{lightgray}{gray}{0.95}
\hypersetup{hidelinks,colorlinks=true,breaklinks=true,urlcolor= blue}
\hypersetup{%
    colorlinks = true,
    linkcolor  = blue,
    citecolor = cyan,
  }
\usepackage{siunitx}
\definecolor{headergray}{gray}{0.90}

\usepackage[font=small,labelfont=bf,labelsep=space]{caption}
\usepackage{natbib}
\setcitestyle{square,numbers}

\newcommand\orcidroldao{{\href{https://orcid.org/0000-0003-3978-532X}{\orcidicon}}}

\newcommand\orcidcasadio{{\href{https://orcid.org/0000-0002-1330-7787}{\orcidicon}}}

\newcommand\orcidandrea{{\href{https://orcid.org/0000-0003-0329-2726}{\orcidicon}}}

\newcommand{\orcidicon}{%
	\begin{tikzpicture}
	\draw[lime, fill=lime] (0,0)
		circle [radius=0.16]
		node[white] {{\fontfamily{qag}\selectfont \tiny ID}};
	\draw[white, fill=white] (-0.0625,0.095)
		circle [radius=0.007];
	\end{tikzpicture}	\hspace{-2mm}
}
\begin{document}
\title{Scalar-tensor reconstruction of coherent quantum black holes}
\author{R.~{Casadio}\orcidcasadio}
\email{casadio@bo.infn.it}
\affiliation{Dipartimento di Fisica e Astronomia, Universit\`a di Bologna, via Irnerio~46, 40126 Bologna, Italy}
\affiliation{I.N.F.N., Sezione di Bologna, I.S.~FLAG, viale B.~Pichat~6/2, 40127 Bologna, Italy}
\affiliation{Alma Mater Research Center on Applied Mathematics (AM\,$^2$), Via Saragozza 8, 40123 Bologna, Italy}
\author{A. {Giusti}\orcidandrea}
\email{andrea.giusti9@unibo.it}
\affiliation{Dipartimento di Fisica e Astronomia, Universit\`a di Bologna, via Irnerio~46, 40126 Bologna, Italy}
\affiliation{I.N.F.N., Sezione di Bologna, I.S.~FLAG, viale B.~Pichat~6/2, 40127 Bologna, Italy}
\affiliation{Alma Mater Research Center on Applied Mathematics (AM\,$^2$), Via Saragozza 8, 40123 Bologna, Italy}
\author{R. {da Rocha}\orcidroldao\!\!}
\email{roldao.rocha@ufabc.edu.br}
\affiliation{Federal University of ABC, Center of Mathematics, Santo Andr\'e, S\~ao Paulo, 09580-210, Brazil}
\begin{abstract}
We construct a scalar-tensor realisation of spherically symmetric black hole geometries
which effectively arise from quantum coherent states characterised by a finite core and finite extension.
We determine the non-minimal coupling to the Einstein--Hilbert Lagrangian, kinetic coupling, and scalar
potential and show that the kinetic term is positive only in finite domains.
For fixed black hole mass, the boundaries of these domains provide a natural mechanism that
relates the core size and spatial extension to the dynamics of the scalar-tensor realisation.
We further show that the same geometry can also be obtained from a Horndeski action with 
derivative coupling and determine its domain of validity.
Neither reconstruction includes the event horizon nor can simultaneously reproduce arbitrarily large portions
of the interior and exterior of the black hole, whose quantum origin therefore appears to remain partially
outside the regime of applicability of semiclassical theories of gravity.
\end{abstract}
\maketitle
\section{Introduction}
\label{sec1}
Understanding the structure of spacetime at scales where classical General Relativity (GR) is expected
to receive significant quantum-gravitational corrections remains one of the central challenges of
gravitational physics.
Although GR provides an exceptionally accurate account of gravity at macroscopic distances,
black hole solutions of the Einstein equations are affected by curvature singularities and geodesic
incompleteness~\cite{Hawking:1973uf}, indicating the limits of the classical theory.
Black holes therefore offer a natural arena for probing such limitations, since their high-curvature regions
are expected to reveal phenomena where quantum-gravitational effects can no longer be neglected.

This problem can be addressed in the corpuscular model~\cite{Dvali:2011aa,Dvali:2012rt}
and the coherent-state approach to gravity~\cite{Casadio:2017cdv,Casadio:2021eio,Muck:2016stv},
in which semiclassical geometries emerge~\cite{Flassig:2012re} collectively from quantum states with large
occupation numbers.
The macroscopic gravitational field generated by a mass $M$ can therefore be described by a coherent state
of $N\sim M^2/\mpl^2$ ``soft'' (non-propagating) gravitons,~\footnote{Hereon, $\mpl$ and $\lp$ denote the
Planck mass and length, $\gn=\lp/\mpl$ the Newton constant and $\hbar=\lp\,\mpl$ the Planck constant.}
whose mean-field reproduces the corresponding spacetime geometry~\cite{Casadio:2015bna,Giusti:2019wdx}.
Finite-$N$ effects then generate corrections suppressed by inverse powers of the occupation number,
which become progressively more relevant for smaller black holes and toward regions where the semiclassical
approximation is expected to break down~\cite{Casadio:2016zpl}.
For instance, entropy and the Hawking evaporation acquire a many-body interpretation in this framework,
with finite-$N$ effects accounting for deviations from exact thermality and the emergence of quantum
hair~\cite{Dvali:2012rt}.
Moreover, the coherent-state approach provides an effective realisation of macroscopic geometries
in which the consistency of the underlying quantum state requires a matter source
of finite size~\cite{Casadio:2021eio} (as well as a finite external
volume~\cite{Meert:2025cct,Bambagiottiab:2026tnr}). 
Related quantum-field-theoretical treatments of graviton bound states have been developed in
Refs.~\cite{Hofmann:2014jya,Casadio:2013hja,Fernandes-Silva:2019fez,Battista:2023iyu}.

These developments have motivated the study of a general class of effective quantum geometries in which
the point-like classical sources are replaced by spatially extended matter
distributions~\cite{Casadio:2015jha,Cadoni:2018dnd}.
Within this context, the relation between quantum hair and information entropy has been investigated
in Refs.~\cite{Feng:2024bsx,Belfiglio:2024qsa,Casadio:2016aum}, while corrections arising during the
gravitational collapse were considered in Ref.~\cite{Calmet:2023met}.
Further extensions include configurations supported by quantum matter cores~\cite{Casadio:2024lzd,Casadio:2023pmh,Feng:2025nai,Casadio:2022pla},
electrically charged coherent black holes~\cite{Casadio:2022ndh,Antonelli:2025mcv},
and, more recently, coherent hairy solutions obtained through the gravitational decoupling~\cite{Navarro:2026lrf}. 
An important consequence of the emergence of an intrinsic length scale associated with the distribution
of the constituent gravitons is that the resulting modification of the short-distance gravitational field
can soften~\cite{Casadio:2021eio,Casadio:2023iqt} or remove curvature divergences~\cite{Urmanov:2024qai},
while Gaussian and related profiles yield regular effective geometries with modified horizon
structures~\cite{Cirilo-Lombardo:2023cxb}.

The modified Schwarzschild geometry consistently emerging from a quantum coherent state requires a 
matter core of finite size, but the shape of this core could be determined only by studying the collapse leading
to the black hole configuration.
Without knowing the full dynamical process, one can either introduce a sharp ultraviolet cut-off in the graviton
distribution~\cite{Casadio:2021eio} or employ a smooth Gaussian profile with a smearing
length scale~\cite{Feng:2025nai,Neznamov:2023hpi}. 
To determine whether this quantum-inspired geometry can arise as an exact solution of a covariant theory,
in this work we embed it in a non-minimally coupled scalar-tensor framework~\cite{Damour:1993hw,Faraoni:2008dn,Peng:2026qnz}
and reconstruct the coupling functions from the prescribed metric and scalar profile.
Requiring regularity and a positive-definite kinetic sector restricts the domain of the areal radius
in which the reconstructed theory remains physically admissible.
This can introduce finite boundaries representing an intrinsic dynamical cutoff of the theory.
As the ratio between the black hole mass and the Gaussian smearing scale is varied,
the topology of this domain can change through the appearance of additional boundaries
or disconnected viable regions.
Such transitions are characteristic of bifurcation phenomena in
relativistic field theories~\cite{Avelino:2008mh,Bandeira:2026rou,Doneva:2022ewd}.
The coherent-state approach thus provides both a quantum-corrected Schwarzschild black hole background
without curvature singularities and a setting in which the domain of validity of its scalar-tensor embedding
can be determined from consistency conditions.

In this work, we show that the non-minimally coupled scalar-tensor reconstruction admits finite radial regions
with a positive Einstein-frame kinetic coefficient.
This establishes a kinetically determined outer radial cutoff beyond which the reconstructed scalar
sector ceases to be admissible, without imposing an external arbitrary cutoff.
We identify a non-degenerate fold bifurcation at which a double zero of the Einstein-frame kinetic
coefficient splits into two simple zeros, generating an additional kinetically excluded radial interval
and potentially disconnecting the admissible domain.
In addition, we implement an exact reconstruction within a derivative-coupled Horndeski theory,
where the Einstein-tensor coupling removes the obstruction in the minimally coupled sector.
In this case, the reconstruction reduces to a linear second-order differential equation driven
by the Gaussian modification of the Schwarzschild geometry.
The resulting kinetic operator determines the locally admissible radial regions through two
independent positivity conditions.
In both frameworks, local admissibility is governed by kinetic consistency, although the resulting
radial domains need not exhibit the same restrictions and differ significantly.

This paper is organised as follows:
in Sec.~\ref{sec2}, we briefly review the coherent-state geometry and formulate its embedding
in a non-minimally coupled scalar-tensor theory;
in Sec.~\ref{sec3}, we implement the inverse reconstruction of the coupling functions and
examine the resulting Einstein-frame kinetic sector.
We determine the physically admissible branches, characterise their finite radial boundaries
and asymptotic behaviour, and analyse the bifurcation structure of the positive-kinetic domains.
The canonically normalised Einstein-frame scalar is then constructed on each connected branch,
establishing its relation with the characteristic scales of the underlying geometry;
in Sec.~\ref{sec4}, we construct an alternative exact realisation within derivative-coupled
Horndeski theory through an Einstein-tensor derivative coupling.
The reconstruction reduces to a linear second-order boundary-value problem,
while the resulting anisotropic kinetic operator determines the corresponding
fixed-background admissible domain;
we conclude by summarising the main results and discussing their physical
implications in Sec.~\ref{sec5}.

\section{Coherent-state geometry and scalar-tensor action}
\label{sec2}

We will first review briefly how a quantum-modified Schwarzschild geometry can emerge from
the expectation value on a suitably defined coherent state~\cite{Casadio:2021eio}.
Consistency of this framework requires that the idealised point-like source of the classical geometry
be replaced by a quantum core of finite size, and the external volume be also finite.
In particular, the length scale associated with the core size will be shown to play a central role
in the reconstruction of an effective field theory of gravity that admits the same modified
Schwarzschild geometry.

\subsection{Semiclassical Schwarzschild geometry}

In order to define a quantum state that can reproduce the Schwarzschild metric
\begin{equation}
\dd s^2
=
-f(r)\,\dd t^2
+\frac{\dd r^2}{f(r)}
+r^2\dd\Omega^2
\ ,
\label{line-element}
\end{equation}
where
\begin{equation}
f=1+2\,V_{\rm N}
=
1-\frac{2\,\gn\,M}{r}
\ ,
\label{V_N}
\end{equation}
one introduces a canonically normalised massless scalar field $\Phi$ satisfying
the field equation
\begin{equation}
\Box\Phi
=
\left[
-\partial_t^2
+
\frac{1}{r^2}\partial_r
\left(r^2\partial_r\right)
\right]\Phi(t,r)
=
0
\ ,
\label{KG-coherent}
\end{equation}
to describe spherically symmetric deviations from the flat Minkowski geometry representing the vacuum
of quantum gravity.
Given the general line element~\eqref{line-element}, we therefore identify the function 
\begin{equation}
V(r)
=
\frac12\left[f(r)-1\right]
\ ,
\label{vp1}
\end{equation}
that must be reproduced by a quantum state of the field $\Phi$.
\par
A positive-frequency basis of solutions to Eq.~\eqref{KG-coherent} is given by
\begin{equation}
u_k(t,r)
=
e^{-i\,k\,t}\,j_0(kr)
\ ,
\label{radial-modes}
\end{equation}
where $j_0(kr)={\sin(kr)}/{kr}$ is the spherical Bessel function of the first kind of order zero.
Hence the quantum field can be expanded as
\begin{equation}
\hat{\Phi}(t,r)
=
\int_0^\infty
\frac{k^2\,\dd k}{2\pi^2}
\sqrt{\frac{\hbar}{2k}}
\left[
u_k(t,r)\hat a(k)
+
u_k^\ast(t,r)\hat a^\dagger(k)
\right]
\ .
\label{Phi-operator}
\end{equation}
The target gravitational configuration can then be obtained from a coherent distribution
of these radial modes, without introducing a classical source explicitly.
Denoting the corresponding momentum-space profile by $g_k$, this coherent state can be
written as~\cite{Feng:2024bsx}
\begin{equation}
\ket{V_M}
=
e^{-N_M/2}\,
\exp\!\left[
\int_0^\infty
\frac{k^2\,\dd k}{2\pi^2}\,
g_k\,\hat a^\dagger(k)
\right]
\ket{0}
\ ,
\label{coherent-state}
\end{equation}
where $N_M$ formally represents a total occupation number which can
be thought of measuring the distance from the Minkowski vacuum $\ket{0}$.
Taking into account the polar decomposition $g_k=|g_k|e^{i\gamma_k},$ 
its mean field reads 
\begin{equation}
\bra{V_M}\hat{\Phi}(t,r)\ket{V_M}
=
\int_0^\infty
\frac{k^2\,\dd k}{2\pi^2}
\sqrt{\frac{2\hbar}{k}}\,
|g_k|
\cos(kt-\gamma_k)
j_0(kr)
\ .
\label{Phi-one-point}
\end{equation}
A static geometry corresponds to phases $\gamma_k=k\,t$ that eliminate the time dependence.

To finally reproduce the Schwarzschild geometry~\eqref{V_N}, we should have
\begin{equation}
\sqrt{\gn}\,
\bra{V_M}\hat{\Phi}(r)\ket{V_M}
=
V_{\rm N}
\ ,
\label{Schwarzschild-potential}
\end{equation}
from which the Fourier--Bessel decomposition yields~\cite{Casadio:2021eio}
\begin{equation}
g_k
=
-\frac{4\pi M}
{\sqrt{2k^3}\,\mpl}
\ .
\label{gk-Schwarzschild}
\end{equation}
Therefore, the momentum distribution associated with the classical Schwarzschild 
solution has the characteristic behaviour $g_k\sim k^{-3/2}$.
However, Eq.~\eqref{gk-Schwarzschild} then gives a total occupation number
\begin{equation}
N_M
=
4\frac{M^2}{\mpl^2}
\int_0^\infty
\frac{\dd k}{k}
\ ,
\label{NM-unregulated}
\end{equation}
which diverges logarithmically both in the infrared and ultraviolet limits of the momentum spectrum.
The low-momentum divergence is associated with the unbounded spatial extent of the classical
Schwarzschild spacetime and can be regulated by introducing a macroscopic length $R_\infty$.
The ultraviolet limit probes arbitrarily short wavelengths, hence the high-momentum divergence is a
consequence of allowing for the source to be infinitesimally small.

A properly normalised coherent state with $N_M<\infty$ is obtained by modifying the short-wavelength
sector of the spectrum~\eqref{gk-Schwarzschild}.
Following Ref.~\cite{Feng:2025nai}, we consider
\begin{equation}
g_k
=
-\frac{4\pi M}
{\sqrt{2k^3}\,\mpl}\,
e^{-k^2\,R_s^2/4}
\ ,
\label{coherent-spectrum}
\end{equation}
where $R_s\ll R_\infty$ characterises the shortest spatial scale resolved by the
coherent gravitational configuration and can be associated with the finite size of a matter core.
Modes with $kR_s\gg1$ are consequently suppressed, whereas the long-wavelength
spectrum is left unchanged.
The total occupation number is thus given by 
\begin{equation}
N_M
=
4\frac{M^2}{\mpl^2}
\int_{R_\infty^{-1}}^\infty
\frac{\dd k}{k}\,
e^{-k^2R_s^2/2}
\ ,
\label{NM-regulated}
\end{equation}
and, for finite $R_s$ and $R_\infty$, both the ultraviolet and infrared regimes are regularised.

The geometric effect of this modification can be seen most directly by transforming
back to position space.
The resulting mean value reads
\begin{equation}
V_{\rm q}(r)
=
-\frac{2\gn M}{\pi r}
\int_0^\infty
\frac{\dd k}{k}\,
e^{-k^2R_s^2/4}\sin(kr)
=
-\frac{\gn M}{r}
\operatorname{erf}\!\left(\frac{r}{R_s}\right)
\ ,
\label{Vq-integral}
\end{equation}
where $R^{-1}_{\infty}$ was let to zero (for simplicity). 
Hence, the momentum-space Gaussian regulator in Eq.~\eqref{coherent-spectrum} 
replaces the singular $V_{\rm N}\sim 1/r$ with a regular $V_{\rm q}$ that is resolved
for $r\sim R_s$.
In this quantum-modified geometry~\cite{Feng:2025nai}
\begin{equation}
f(r)
=
1+2\,V_{\rm q}(r)
=
1-\frac{2\gn M}{r}
\operatorname{erf}\!\left(\frac{r}{R_s}\right)
\ ,
\label{metric}
\end{equation}
the location of the event horizon $r=r_h$ is still determined by $f(r_h)=0$ or
\begin{equation}
r_h
=
2\gn M\,\operatorname{erf}\!\left(\frac{r_h}{R_s}\right)
\ ,
\label{rh}
\end{equation}
so that $r_h>0$ only for $R_s<4 \gn M/\sqrt{\pi}\simeq 1.13\,(2 \gn M)$.

\subsection{Effective scalar-tensor theory}

We now want to reconstruct a scalar-tensor action admitting the metric~\eqref{metric}
as an exact solution.
We remark that the scale $R_s$ plays a twofold role:
it regulates the high-momentum sector of the coherent-state spectrum and sets the characteristic
radial scale below which the geometry deviates from the classical Schwarzschild solution.
In the asymptotic regime $r/R_s\gg1$, the error function approaches unity, and the classical
Schwarzschild metric is recovered.
For $r\lesssim R_s$, finite-resolution effects dominate and induce corresponding modifications
to the geometry.

The Jordan-frame action must be of the form
\begin{eqnarray}
\label{act1}
S_J
&=& \frac{1}{16\pi \gn}
\int \dd^4x\,\sqrt{-g}\,
\left[
F(\phi)\,\mathcal R
-
Z(\phi)\,\nabla_\mu\phi\,\nabla^\mu\phi
-
2U(\phi)
\right],
\end{eqnarray}
where $F(\phi)$ describes the non-minimal coupling between the scalar field $\phi$
and the Ricci scalar $\mathcal R$, and $U$ is a potential for $\phi$.~\footnote{The minimal coupling
$F(\phi)\equiv 1$ is not viable for actions of the form~\eqref{act1}.
In fact, the Einstein tensor satisfies $G^t_{\ t}=G^r_{\ r}$ for the line element~\eqref{line-element} and, if $F(\phi)=1$,
subtracting the corresponding field equations gives $Z(\phi)f(r)[\phi^{\prime}(r)]^2=0$.
Since $\phi'(r)\neq0$ for the profile~\eqref{phi1}, one obtains $Z(\phi)=0$ wherever $f(r)\neq0$.
The scalar field equation then implies $U_{,\phi}=0$, whereas the $tt$--Einstein equation
for the metric~\eqref{metric} would require
\begin{eqnarray}
U(r)=\frac{4\gn M}{\sqrt{\pi}R_s r^2}
\exp\!\left(-\frac{r^2}{R_s^2}\right)
\ ,
\end{eqnarray}
which is not constant.
Equivalently, the angular Einstein equation is incompatible with the prescribed metric,
and the geometry~\eqref{metric} \emph{requires} a non-minimal coupling $F(\phi)\neq 1$
for its reconstruction.
Moreover, this obstruction is independent of the specific scalar profile~\eqref{phi1}
and applies to any non-constant monotonic configurations, such as $\phi(r)=\tanh(r/R_s)$
or $\phi(r)=1-\exp(-r^2/R_s^2)$.
\label{foot2}}
In particular, we will consider 
\begin{equation}
\label{phi1}
\phi(r)={\rm erf}\left(\frac{r}{R_s}\right)
\ ,
\end{equation}
and primes will denote total radial derivatives after evaluation on the background~\eqref{phi1}.
For example, 
\begin{equation}
F'
\equiv
\frac{\dd}{\dd r}F(\phi(r))
=
F_{,\phi}\,\phi'
\ ,
\end{equation}
where $F_{,\phi}$ denotes the derivative with respect to the scalar field.

Varying the action~\eqref{act1} with respect to the metric yields the field equations 
\begin{equation}
F \,G_{\mu\nu}
=
Z\left(
\nabla_\mu\phi\nabla_\nu\phi
-\frac12 g_{\mu\nu}
\nabla_\alpha\phi\nabla^\alpha\phi
\right)
-Ug_{\mu\nu}
+\nabla_\mu\nabla_\nu F
-g_{\mu\nu}\Box F
\ ,
\label{metric-eq}
\end{equation}
where $G_{\mu\nu}$ is the Einstein tensor.
Likewise, variation of the action~\eqref{act1} with respect to the scalar field yields 
\begin{equation}\label{te1}
F_{,\phi}\,\mathcal R
+
2Z\,\Box\phi
+
Z_{,\phi}\nabla_\mu\phi\nabla^\mu\phi
-
2U_{,\phi}
=
0
\ .
\end{equation}
For the scalar profile~\eqref{phi1}, one has
\begin{equation}
F_{,\phi}
=
\frac{F'}{\phi'}
=
\frac{\sqrt{\pi}R_s}{2}
e^{r^2/R_s^2}F'
\ .
\end{equation}
The contracted Bianchi identity (equivalent to diffeomorphism invariance) then implies
that Eq.~\eqref{te1} follows from the covariant divergence of Eq.~\eqref{metric-eq}.

Spherical symmetry leaves three independent components of the Einstein tensor,
which can be conveniently expressed as
\begin{subequations}
\begin{eqnarray}
G^t_{\ t}
&=&
-\frac{r f'+f^2-f}{r^2f^2}
\ ,
\label{Gtt}
\\
G^r_{\ r}
&=&
\frac{r f'-f^2+f}{r^2f^2}
\ ,
\label{Grr}
\\
G^\theta_{\ \theta}
=
G^\varphi_{\ \varphi}
&=&
\frac{1}{2f^2}
\left(
f''-\frac{f^{\prime 2}}{f}
\right)
\ .
\label{Gangular}
\end{eqnarray}
\end{subequations}
The Ricci scalar then reads 
\begin{equation}
\mathcal R
=
-\frac{f''}{f^2}
+\frac{f^{\prime 2}}{f^3}
+\frac{2}{r^2}
-\frac{2}{r^2f}
\ .
\label{Ricci}
\end{equation}
Evaluating all terms for the profile~\eqref{phi1}, the metric field equations~\eqref{metric-eq} provide
three independent relations that determine the coupling functions in the action~\eqref{act1}.
In particular, subtracting the $tt$ component from the $rr$ component eliminates the potential $U$ and yields
\begin{equation}
Z\!\left(\phi(r)\right)
=
\frac{\pi R_s^2}{4}
e^{2r^2/R_s^2}
\left[
-F''
+
\frac{f'}{f}
\left(
F'+\frac{2F}{r}
\right)
\right]
\ .
\label{Zcompact}
\end{equation}
Substituting Eq.~\eqref{Zcompact} into the $tt$-component of
Eq.~\eqref{metric-eq} then determines the scalar potential as
\begin{equation}
U\!\left(\phi(r)\right)
=
-\frac{1}{2f}
\left[
F''
+\frac{4F'}{r}
+\frac{2(f-1)}{r^2}F
\right]
\ .
\label{Ucompact}
\end{equation}
After eliminating $Z$ and $U$ through the $tt$ and $rr$ equations,
the remaining independent angular equation reduces to a
first-order homogeneous equation for the non-minimal coupling,
\begin{equation}
\frac{F'}{F}
=
-\frac{
r^2\!\left(ff''-f^{\prime 2}\right)
+2rff'
+2f^2(f-1)
}{
rf\left(rf'-2f\right)}
\ .
\label{logFcompact}
\end{equation}
which determines $F$ where its denominator does not vanish.
Once $F$ is calculated, the kinetic coupling $Z$ and the potential $U$ follow directly
from Eqs.~\eqref{Zcompact} and \eqref{Ucompact}, respectively.
\par 
From Eqs.~\eqref{metric} and~\eqref{logFcompact}, integrating on a connected regular branch
yields
\begin{equation}
F\!\left({\rm erf}\left(\frac{r}{R_s}\right)\right)
=
F_0
\exp\!\left[
-\int_{r_0}^{r}
\frac{\mathcal N(s)}{\mathcal D(s)}\,\dd s
\right]
\ ,
\label{Fexplicit}
\end{equation}
where $r_0$ is an arbitrary point chosen within the same connected regular branch
of $r$, and
\begin{equation}
F_0
\equiv
F\!\left({\rm erf}\left(\frac{r_0}{R_s}\right)\right)
\end{equation}
is the corresponding normalisation constant.
The numerator and denominator in the integral in Eq.~\eqref{Fexplicit} read
\begin{align}
\mathcal N(s)
=
\frac{4\gn M\,s^2}{R_s^3}
&
\left[
\frac{2}{\sqrt\pi} e^{-s^2/R_s^2}
-
\frac{4\,\gn M R_s}{\pi\,s^2} e^{-2s^2/R_s^2}
+
\frac{4\,\gn M}{\sqrt\pi s} e^{-s^2/R_s^2}
\left(\frac{R_s^2}{s^2}-1\right)
\operatorname{erf}\!\left(\frac{s}{R_s}\right)
\right.
\nonumber
\\
&
\ \
\left.
-
\frac{R_s^3}{s^3}
\operatorname{erf}\!\left(\frac{s}{R_s}\right)
+
\frac{3\gn M R_s^3}{s^4}
\operatorname{erf}^2\!\left(\frac{s}{R_s}\right)
-
\frac{4\gn^2M^2R_s^3}{s^5}
\operatorname{erf}^3\!\left(\frac{s}{R_s}\right)
\right]
\ ,
\label{Nexplicit}
\end{align}
and
\begin{equation}
\mathcal D(s)
=
2\,s
\left[
\frac{2\gn M}{s}\operatorname{erf}\!\left(\frac{s}{R_s}\right)
-1
\right]
\left[
1
+
\frac{2\gn M}{\sqrt{\pi}\,R_s} e^{-s^2/R_s^2}
-
\frac{3\gn M}{s}
\operatorname{erf}\!\left(\frac{s}{R_s}\right)
\right]
\ .
\quad
\label{Dexplicit}
\end{equation}
We will analyse in details the above reconstruction in the next Section.

\section{Scalar-tensor kinetic sector and radial viability}
\label{sec3}

The above reconstruction ensures that the metric~\eqref{metric} and the scalar field~\eqref{phi1}
satisfy the background field equations, but does not guarantee that the action~\eqref{act1} from
which such equations are derived is dynamically consistent.
We now address the physical viability of the reconstructed scalar-tensor theory by requiring
a positive effective gravitational coupling and a positive scalar kinetic term in the Einstein frame.
\par
The two frames are related by the conformal transformation
\begin{equation}
g_{\mu\nu}^{E}
=
F(\phi)\,g_{\mu\nu}^{J}
\ ,
\label{Einsteintransform}
\end{equation}
where $E$ and $J$ label quantities in the Einstein and Jordan frames, respectively.
We first require that $0<F(\phi)<\infty$, for finite values of $\phi$, to ensure that the conformal
transformation is nonsingular and that the Einstein--Hilbert term has the standard sign.
The action~\eqref{act1} can therefore be rewritten, up to the standard boundary term, as
\begin{equation}
S_E
=
\frac{1}{16\pi \gn}
\int \dd^4x\,\sqrt{-g_E}
\left[
\mathcal R_E
-
K_E(\phi)
\nabla_\mu^E\phi\,\nabla_E^\mu\phi
-
2V_E(\phi)
\right]
\ ,
\label{Einsteinaction}
\end{equation}
where
\begin{equation}
V_E(\phi)
=
\frac{U(\phi)}{F(\phi)^2}
\label{VE}
\end{equation}
and
\begin{equation}
K_E(\phi)
=
\frac{Z(\phi)}{F(\phi)}
+
\frac{3}{2}
\left(
\frac{F_{,\phi}}{F}
\right)^2
\ ,
\label{KEdefinition}
\end{equation}
so that the scalar kinetic term has the correct sign for $K_E(\phi)>0$.

Since the normalisation~\eqref{NM-regulated} of the coherent state requires $R_s>0$,
it is convenient to introduce the dimensionless variables~\footnote{In the limit $R_s\to 0$,
Eq.~\eqref{metric} reduces to the exact Schwarzschild metric which is obtained from
the action~\eqref{Einsteinaction} with $K_E=V_E=0$.}
\begin{equation}
x=\frac{r}{R_s}
\ ,
\qquad
\mu=\frac{\gn M}{R_s}
\ ,
\label{d1}
\end{equation}
so that the metric function in Eq.~\eqref{metric} reads
\begin{equation}
f(x,\mu)
=
1-\frac{2\mu}{x}\operatorname{erf}(x)
=
1-\frac{2\mu}{x}\,\phi(x)
\ .
\label{mx}
\end{equation}
We remark that an event horizon $x_h>0$ where $f(x_h,\mu)=0$ exists only for
\begin{equation}
\mu>\mu_h
=
\frac{\sqrt{\pi}}{4}
\simeq
0.44
\ ,
\label{muh}
\end{equation}
although we will consider all values of $\mu>0$ in the following for completeness.

Denoting the logarithmic derivative with
\begin{equation}
Y(x,\mu)
\equiv
\frac{\dd}{\dd x}\ln F(x),
\label{Ydefinition}
\end{equation}
the Einstein-frame kinetic coefficient evaluated on the background
can be written as~\footnote{From now on, $K_E(x,\mu)$ denotes the Einstein-frame kinetic
coefficient $K_E(\phi)$ evaluated on the background $\phi(x)=\operatorname{erf}(x)$.}
\begin{eqnarray}
K_E(x,\mu)
&=&
\frac{\pi}{4}e^{2x^2}
\left[
-Y_{,x}
+\frac{1}{2}Y^2
+
\frac{f_{,x}}{f}
\left(
Y+\frac{2}{x}
\right)
\right]
\nonumber
\\
&=&
\frac{\pi}{4}e^{2x^2}
\frac{\mathcal P(x,\mu)}
{\mathfrak D(x,\mu)}
\ ,
\label{KEdimensionless}
\end{eqnarray}
where
\begin{eqnarray}
\mathcal P(x,\mu)
&\!\!=\!\!&
32\pi\mu^6\operatorname{erf}^6(x)
-96\pi\mu^5x\operatorname{erf}^5(x)
+66\pi\mu^4x^2\operatorname{erf}^4(x)
\nonumber
\\
&&
+
2\mu^3x^3\operatorname{erf}^3(x)
\left[
-48\sqrt{\pi}\,\mu e^{-x^2}x^4
+120\sqrt{\pi}\,\mu e^{-x^2}x^2
+48\sqrt{\pi}\,\mu e^{-x^2}
+2\pi
\right]
\nonumber
\\
&&
+
2\mu^2x^4\operatorname{erf}^2(x)
\left[
16\mu^2e^{-2x^2}x^4
-256\mu^2e^{-2x^2}x^2
+72\mu^2e^{-2x^2}
\right.
\nonumber
\\
&&
\hspace{2.8cm}
\left.
+\,64\sqrt{\pi}\,\mu e^{-x^2}x^4
-116\sqrt{\pi}\,\mu e^{-x^2}x^2
-108\sqrt{\pi}\,\mu e^{-x^2}
-3\pi
\right]
\nonumber
\\
&&
+
8\mu^2x^5\operatorname{erf}(x)
\left[
\frac{24}{\sqrt{\pi}}\mu^2e^{-3x^2}x^2
-\frac{48}{\sqrt{\pi}}\mu^2e^{-3x^2}
-4\mu e^{-2x^2}x^4
\right.
\nonumber
\\
&&
\left.
\phantom{\frac{A}{B}}
\hspace{2.2cm}
+52\mu e^{-2x^2}x^2
+18\mu e^{-2x^2}
-7\sqrt{\pi}\,e^{-x^2}x^4
+8\sqrt{\pi}\,e^{-x^2}x^2
+9\sqrt{\pi}\,e^{-x^2}
\right]
\nonumber
\\
&&
+
\frac{160}{\pi}\mu^4e^{-4x^2}x^6
-\frac{32}{\sqrt{\pi}}\mu^3e^{-3x^2}x^6
\left(3x^2-1\right)+
8\mu^2e^{-2x^2}x^6
\left(x^4-10x^2-6\right)
\nonumber
\\
&&
+
4\sqrt{\pi}\,\mu e^{-x^2}x^6
\left(2x^4-x^2-1\right)
\label{PKE}
\end{eqnarray}
and
\begin{equation}
\label{dx1}
\mathfrak{D}(x,\mu)
=
x^4
\left[x-2\mu\operatorname{erf}(x)\right]^2
\left[
\sqrt{\pi}\,x
+2\mu x e^{-x^2}
-3\sqrt{\pi}\,\mu\operatorname{erf}(x)
\right]^2
\ .
\end{equation}
Eq.~\eqref{KEdimensionless} therefore provides a closed expression for the Einstein-frame
kinetic coefficient in terms of the dimensionless variables~\eqref{d1} and~\eqref{mx}.
\par
In the same dimensionless variables, the denominator~\eqref{Dexplicit} that appears 
in the expression of $F$ in Eq.~\eqref{Fexplicit} reads
\begin{eqnarray}
\mathcal D(x,\mu)
&\!\!=\!\!&
-\frac{2R_s}{\sqrt{\pi}\,x}
\left[x-2\mu\operatorname{erf}(x)\right]
\left[
\sqrt{\pi}\,x
+2\mu x e^{-x^2}
-3\sqrt{\pi}\,\mu\operatorname{erf}(x)
\right]
\nonumber
\\
&=&
\pm\frac{2R_s}{\sqrt{\pi}\,x^3}
\sqrt{\mathfrak D(x,\mu)}
\ ,
\label{dd1}
\end{eqnarray}
and the numerator in Eq.~\eqref{Nexplicit} is given by
\begin{eqnarray}
\mathcal N(x,\mu)
&=&
4\mu x^2
\left[
\frac{2}{\sqrt{\pi}}e^{-x^2}
-\frac{4\mu}{\pi x^2}e^{-2x^2}
+\frac{4\mu}{\sqrt{\pi}x}e^{-x^2}
\left(\frac{1}{x^2}-1\right)
\operatorname{erf}(x)\right.
\nonumber\\
&&\left.\qquad\qquad\qquad\qquad-\frac{1}{x^3}\operatorname{erf}(x)
+\frac{3\mu}{x^4}\operatorname{erf}^2(x)
-\frac{4\mu^2}{x^5}\operatorname{erf}^3(x)
\right]
\ .
\label{Nn1}
\end{eqnarray}
Necessary conditions for the theory to be physically viable are that the above expressions make
$K_E(x,\mu)$ and $F(x,\mu)$ strictly positive and finite for finite $x$ (so that the logarithmic derivative
of the latter, $Y(x,\mu)$, is also finite).

\subsection{Kinetic consistency and reconstruction poles}
\label{subsec1}

The physically admissible domain for the action~\eqref{Einsteinaction} at fixed $\mu$ is defined as
\begin{equation}
\mathcal I(\mu)
=
\left\{
x\in \mathbb{R}^+\,\big|\,
K_E(x,\mu)>0,\;
|Y(x,\mu)| < \infty
\right\},
\label{s1}
\end{equation}
with $F(x,\mu)>0$ understood throughout.
We will show later on that this domain contains connected branches that we denote with $\mathcal I_a(\mu)$.

Regarding the condition $|Y(x,\mu)| < \infty$, we note that the denominator $\mathcal{D}(x,\mu)$
in Eq.~\eqref{dd1} vanishes at points that satisfy at least one of the equalities
\begin{eqnarray}
x
&=&
2\mu\operatorname{erf}(x)
\ ,
\label{Dzero1}
\\
x
&=&
3\,\mu\operatorname{erf}(x)
-\frac{2\mu}{\sqrt{\pi}} x e^{-x^2}
\ .
\label{Dzero2}
\end{eqnarray}
Such zeros identify genuine singularities of the conformal transformation~\eqref{Einsteintransform}
only if they are not canceled by simultaneous zeros of the numerator $\mathcal N(x,\mu)$ in
Eq.~\eqref{Nn1}.
We call {\em reconstruction poles\/} these singularities at fixed $\mu$ and denote them with $x=x_Y(\mu)$.
Such poles mark the boundaries of regular reconstruction branches where the conformal
transformation~\eqref{Einsteintransform} is well-defined.
Since Eqs.~\eqref{Dzero1} and~\eqref{Dzero2} are transcendental, the values of 
$x_Y(\mu)$ can be determined only numerically.

Remarkably, Eq.~\eqref{Dzero1} is precisely the horizon Eq.~\eqref{rh} in dimensionless
variables, which means that $x=x_h=r_h/R_s$ is always going to be a reconstruction pole
unless $\mathcal N(x_h,\mu)$ also vanishes.~\footnote{We recall that the horizon only exists
for $\mu>\mu_h\simeq 0.44$ in Eq.~\eqref{muh}.}
Moreover, since the denominator $\mathcal{D}(x,\mu)\sim \mathfrak{D}(x,\mu)$ in Eq.~\eqref{dx1},
the same zeros of $\mathcal{D}(x,\mu)$ are potential singularities of $K_E(x,\mu)$
in Eq.~\eqref{KEdimensionless} as well, provided they are not cancelled by simultaneous zeros
of $\mathcal{P}(x,\mu)$ in Eq.~\eqref{PKE}.

Inside regular reconstruction branches we must also require that $K_E(x,\mu)>0$.
We call {\em kinetic boundaries\/} the locations $x=x_K(\mu)$ where $K_E(x,\mu)$
at fixed $\mu$ changes sign.
This condition can be studied analytically for $0<x\ll 1$ and $x\gg 1$.

The near-origin behaviour of $K_E(x,\mu)$ can be analysed by expanding the metric
function~\eqref{metric} for $0<x\ll 1$, 
\begin{equation}
f(x)
=
1-\frac{4\mu}{\sqrt{\pi}}
+
\frac{4\mu}{3\sqrt{\pi}}x^2
+
\mathcal O(x^4)
\ ,
\label{fsmallx}
\end{equation}
which, upon substituting into Eq.~\eqref{KEdimensionless}, yields
\begin{equation}
K_E(x,\mu)
=
\frac{\pi}{8x^2}
\left[
\left(
1-\frac{4\mu}{\sqrt{\pi}}
\right)^2-1
\right]
+
\mathcal O(1)
\ .
\label{KEsmallx}
\end{equation}
The above leading-order coefficient vanishes at $\mu=\sqrt{\pi}/2$, 
where the sign of $K_E$ is determined by the next non-vanishing term
in the expansion.
It follows that, for $0<\mu<{\sqrt{\pi}}/{2}$, the kinetic coefficient is negative in a sufficiently
small neighbourhood of the origin, whereas it is positive for $\mu>{\sqrt{\pi}}/{2}$
in such a neighbourhood.
This result implies that the action~\eqref{Einsteinaction} cannot describe coherent
state black holes with a core of arbitrarily small size $R_s\ll \gn M$.

We next examine the asymptotic behaviour $x\gg1$, where Eq.~\eqref{metric} reads
\begin{equation}
f(x)
=
1-\frac{2\mu}{x}
+\mathcal O(e^{-x^2})
\ .
\label{exp1}
\end{equation}
Eqs.~\eqref{Ydefinition} and~\eqref{KEdimensionless} then give
\begin{equation}
K_E(x,\mu)
=
-\frac{\pi}{4}e^{2x^2}
\left[
\frac{6\mu^2}{x^4}
+\frac{56\mu^3}{x^5}
+\frac{272\mu^4}{x^6}
+
\mathcal O(x^{-7})
\right]
\ .
\label{KElargex}
\end{equation}
We conclude that, for fixed $\mu$, there exists a finite radius $x=x_K(\mu)$ beyond
which $K_E(x,\mu)$ is negative (see also Appendix~\ref{app2}).
This result follows directly from the asymptotic structure of the reconstructed theory
and shows that no connected regular branch with $K_E>0$ can extend to spatial infinity.
This feature is compatible with the necessary infrared regularisation of the coherent state
in Eq.~\eqref{NM-regulated}, which would allow us to identify $x_c(\mu)\simeq R_\infty/R_s$.

\subsection{Viable radial intervals}
\label{subsec3}
We will now determine in details the connected components of the admissible radial
domain $\mathcal I(\mu)$ defined in Eq.~\eqref{s1}.
Each connected component at fixed $\mu$ is a region
$\mathcal I_a=\{x\in \mathbb{R}^+\,|\, x_{a-}(\mu)<x<x_{a+}(\mu)\}$,
at whose boundaries either $K_E$ changes sign or $Y$ diverges.

The boundaries $x_\pm=x_{K}(\mu)$ where the scalar kinetic term changes sign are determined by 
\begin{equation}
K_E(x_K,\mu)=0
\ ,
\label{KEzero}
\end{equation}
with $K_{E,x}(x_{K},\mu)\neq 0$.
We will show that there are also solutions of Eq.~\eqref{KEzero} representing
bifurcation points $(x_{\rm bif},\mu_{\rm bif})$ in the two-dimensional space $(x,\mu)$
where $K_{E,x}(x_{\rm bif},\mu_{\rm bif})=0$.

Using Eq.~\eqref{KEdimensionless}, on every regular branch where Eqs.~\eqref{Dzero1}
and~\eqref{Dzero2} are not satisfied, the kinetic boundary condition~\eqref{KEzero}
can be equivalently written as
\begin{equation}
\mathcal P(x,\mu)=0
\ ,
\label{KEPequivalence}
\end{equation}
where $\mathcal P$ is given in Eq.~\eqref{PKE}.
Indeed, differentiating Eq.~\eqref{KEdimensionless} with respect to $x$ gives
\begin{equation}
\lim_{\mathcal P\to0}K_{E,x}
=
\frac{\pi}{4}e^{2x^2}
\frac{\mathcal P_{,x}(x,\mu)}
{x^4
\left[x-2\mu\operatorname{erf}(x)\right]^2
\left[
\sqrt{\pi}\,x
+2\mu x e^{-x^2}
-3\sqrt{\pi}\,\mu\operatorname{erf}(x)
\right]^2}
\ .
\label{KExP}
\end{equation}
Consequently, the bifurcation conditions $K_E(x,\mu)=K_{E,x}(x,\mu)=0$
reduce to
\begin{equation}
{\mathcal P}(x,\mu)
=
{\mathcal P}_{,x}(x,\mu)
=
0
\ ,
\label{Pbifconditions}
\end{equation}
on every regular branch inside $\mathcal I(\mu)$.

Since all equations involve $\operatorname{erf}(x)$ and $e^{-x^2}$,
we expect that there are no closed-form algebraic solutions for the boundaries of $\mathcal I_a$
and bifurcation points. 
We therefore solve Eqs.~\eqref{Pbifconditions} numerically in the dimensionless
domain $0<x<30$ and $0<\mu<100$.~\footnote{The point $x=0$ lies outside the numerical
search domain and is covered by the small-$x$ expansion~\eqref{KEsmallx}.}
The simultaneous roots are obtained in \textit{Mathematica} using \texttt{FindRoot} from a grid of initial
seeds distributed over the numerical search domain, with increased working precision used to verify
convergence. 
The numerical analysis identifies a regular bifurcation
point located at~\footnote{Numerical values
are reported with higher precision at the bifurcation point, while three decimal
places are retained for the boundaries of $\mathcal I_a$ shown in the figures
throughout this work.}
\begin{equation}
\mu_{\rm bif}=1.61347066
\ ,
\qquad
x_{\rm bif}=1.39520393
\ .
\label{bif1}
\end{equation}
In Appendix~\ref{app1}, we show that the critical point~\eqref{bif1}
indeed satisfies the conditions characterising a fold (saddle-node)
bifurcation. No additional regular solutions of Eqs.~\eqref{Pbifconditions} are found
within the numerical search domain at the adopted working precision
(see also Appendix~\ref{app3} for a more general argument).

\begin{figure}[t]
\centering
\includegraphics[width=11cm]{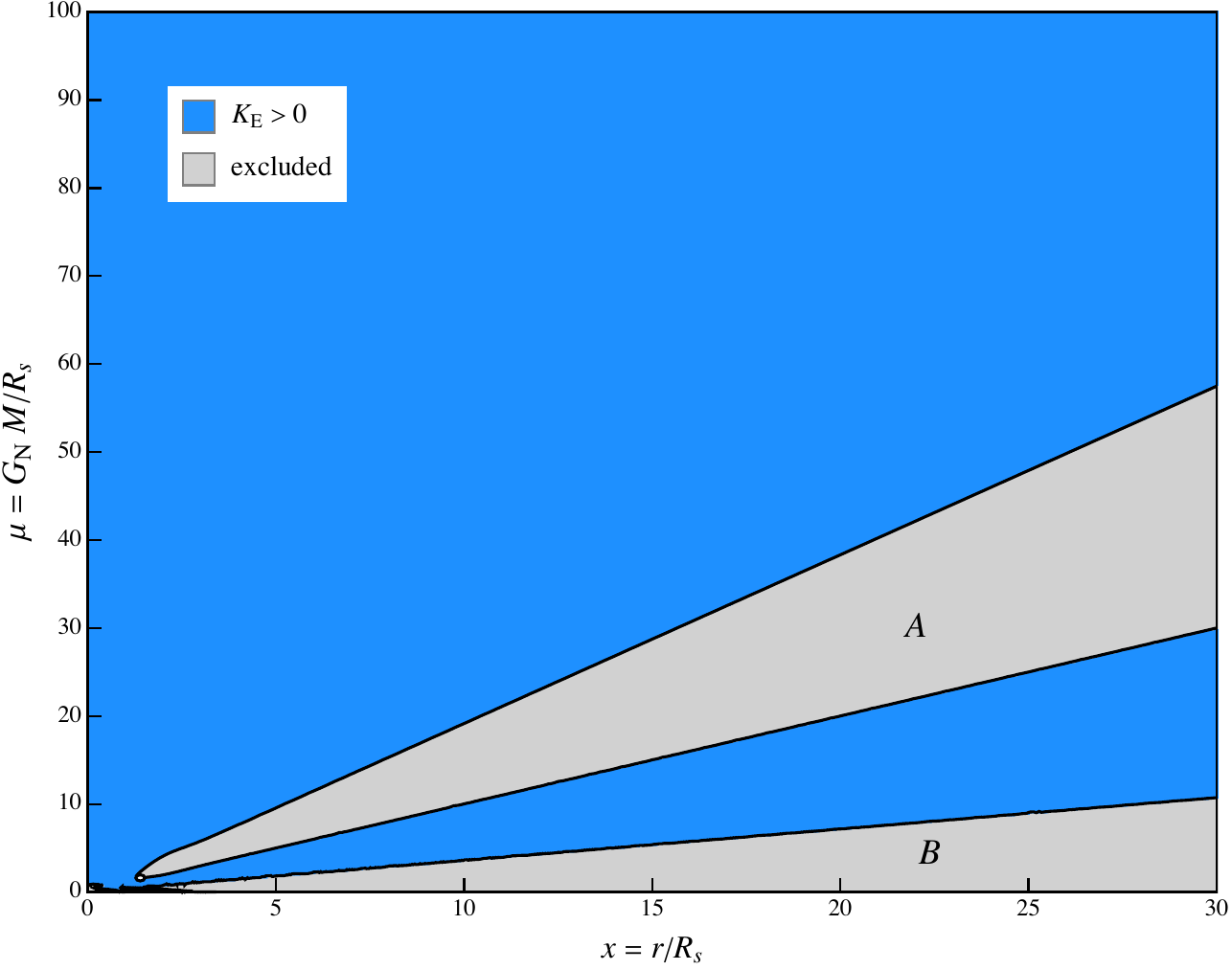}
\caption{Domain of viability for the reconstructed scalar-tensor theory:
$K_E(x,\mu)>0$ inside the blue region and $K_E(x,\mu)<0$ inside the light gray regions.
The black hole event horizon lies slightly above the upper border of the excluded region
$B$ (see Fig.~\ref{plot12}).}
\label{plot1}
\end{figure}

\begin{figure}[t]
\centering
\includegraphics[width=11.5cm]{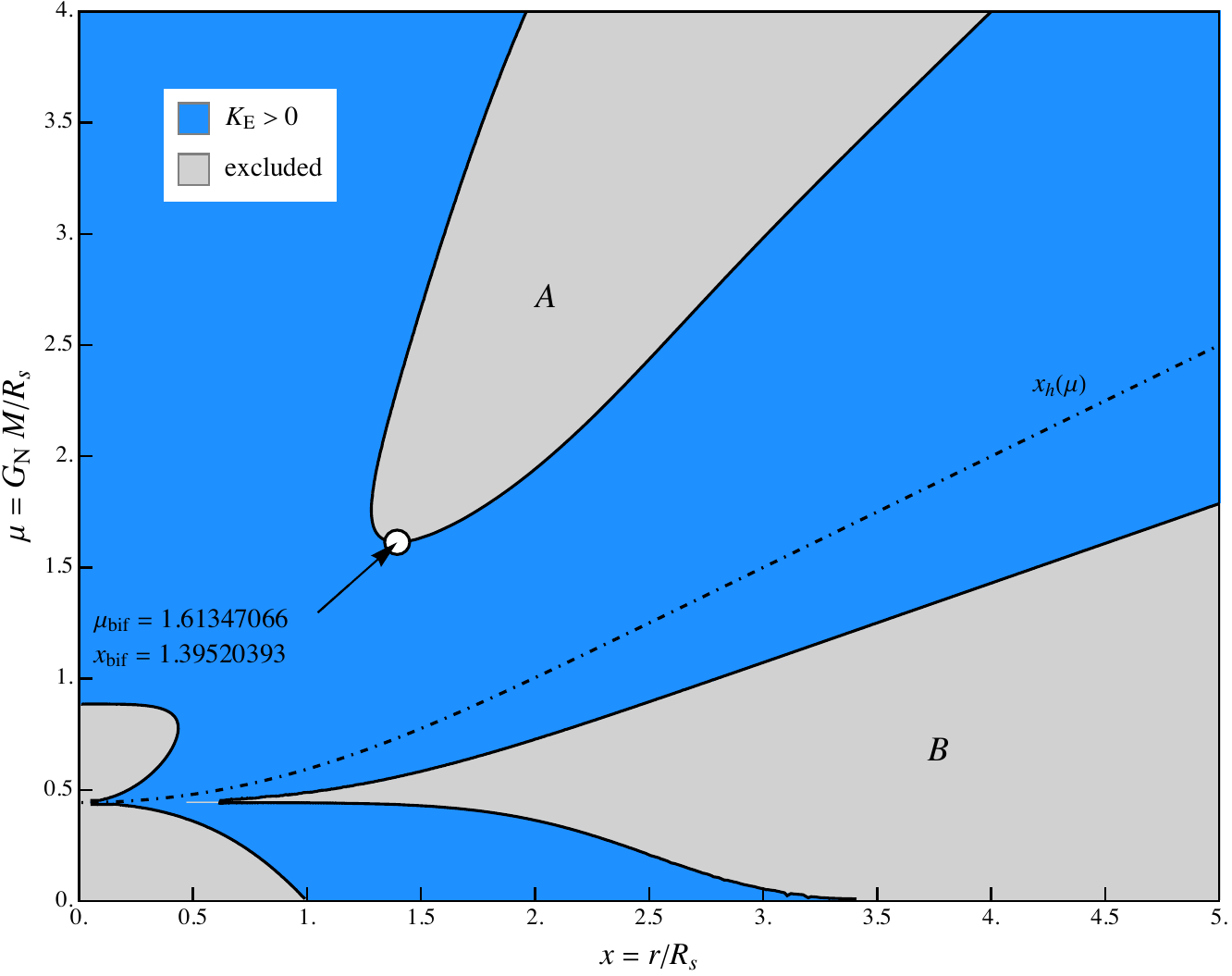}
\caption{Domain of viability for the reconstructed scalar-tensor theory:
$K_E(x,\mu)>0$ inside the blue region and $K_E(x,\mu)<0$ inside the light gray regions.
The event horizon of the coherent-state black hole is located at $x=x_h$ (dashed-dotted line).}
\label{plot12}
\end{figure}

We likewise search for the boundaries of connected regions inside $\mathcal I(\mu)$
numerically.
Fig.~\ref{plot1} shows the values of $(\mu,x)$ for which the kinetic
term is positive and consistent with the reconstruction procedure.
A closer view of the region around the bifurcation point is shown in Fig.~\ref{plot12},
where one the location $x=x_h(\mu)$ of the event horizon is visible (and
remains very close to the upper border of the excluded region $B$).
An important remark is that, for fixed black hole mass $M$ and core radius $R_s$
(hence, for fixed $\mu$), the connected allowed region between the excluded
regions $A$ and $B$ extends from the interior to the exterior, but the size of the latter
is never much larger than $x_h$.

\subsection{Canonical scalar field and kinetic branches}
\label{subsec4}

The condition $K_E>0$ admits a direct field-theoretical interpretation in terms of a canonically
normalised scalar degree of freedom.
On each connected branch $\mathcal I_a\subset\mathcal I(\mu)$,
the canonically normalised Einstein-frame scalar field is defined by
\begin{equation}
\frac{\dd\chi}{\dd\phi}
=
\sqrt{K_E(\phi)}
\ .
\label{cfd1}
\end{equation}
Since the profile~\eqref{phi1} is strictly monotonic for every $x>0$,
this construction can be expressed directly in terms of the radial variable.
Using $\phi_{,x}=2e^{-x^2}/\sqrt{\pi}$ together with Eq.~\eqref{KEdimensionless},
on the $a^{\rm th}$-connected component of the admissible domain $\mathcal I(\mu)$, 
one obtains
\begin{equation}
\chi_a(x)
=
\chi_{0a}
+
\int_{x_{0a}}^{x}
\sqrt{
\frac{\mathcal P(\xi,\mu)}
{\mathfrak D(\xi,\mu)}
}\,\dd\xi
\ ,
\label{chixintegral}
\end{equation}
where $x_{0a}$ is an arbitrary point within the same branch.
The canonical field is therefore defined independently on each connected component
$\mathcal I_a$.

Since $K_E>0$ in the interior of $\mathcal I_a$ and
$\phi_{,x}>0$ for every finite $x$, the mapping 
$x\mapsto\phi(x)$ and $\phi\mapsto\chi_a(\phi)$ are strictly monotonic
on each such interval.
The canonical transformation can therefore be inverted, $\phi=\phi_a(\chi),$ 
in the interior of each connected $\mathcal I_a$.
In terms of the field $\chi$, the Einstein-frame action~\eqref{Einsteinaction}
takes the form
\begin{equation}
S=\frac{1}{16\pi \gn}\int \dd^4x\,\sqrt{-g_E}
\left[
\mathcal R_E-\nabla_\mu^E\chi\,\nabla_E^\mu\chi
-2V_{Ea}(\chi)
\right],
\label{e11}
\end{equation}
where
\begin{equation}
V_{Ea}(\chi)
=
\left.
\frac{U(\phi)}{F(\phi)^2}
\right|_{\phi=\phi_a(\chi)}
\ .
\label{canonical-potential}
\end{equation}
Thus, every connected component $\mathcal I_a$ of the positive-kinetic domain 
$\mathcal I(\mu)$ is associated with a canonical Einstein-frame scalar together
with a corresponding reconstructed potential.

\begin{figure}[t]
\centering
\includegraphics[width=11cm]{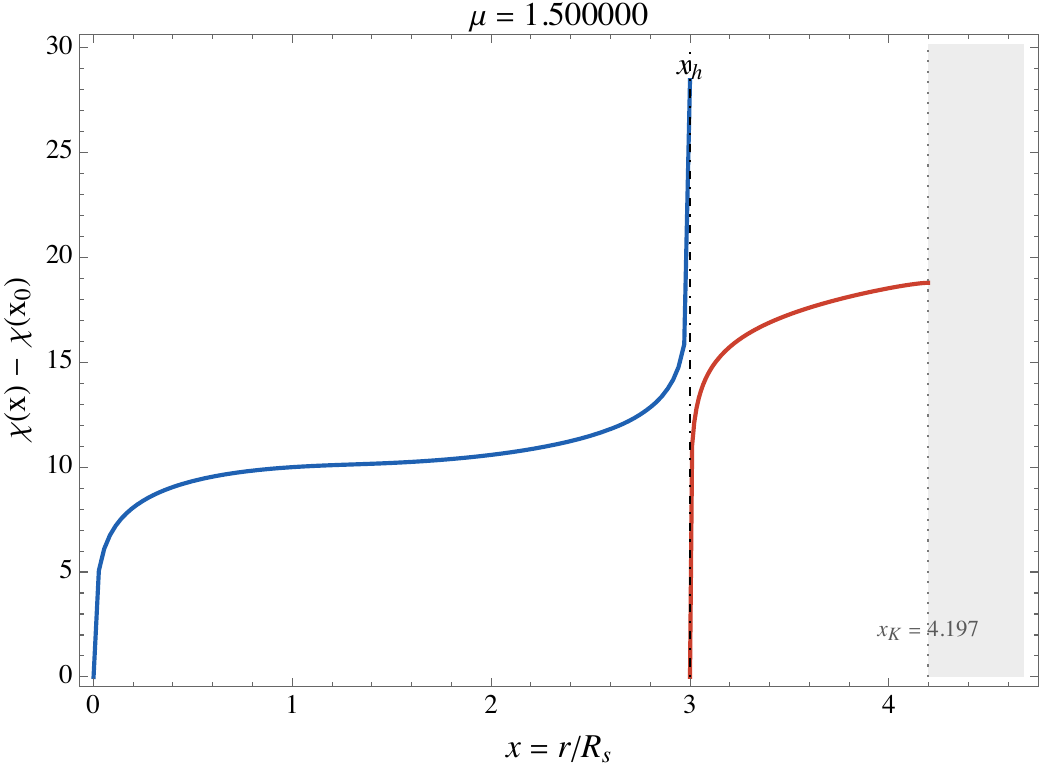}
\caption{Canonical scalar field $\chi(x)$ for $\mu=1.5<\mu_{\rm bif}$.
The dash-dotted vertical line indicates the reconstruction pole $x_Y=x_h$,
whereas the dotted vertical line marks the
boundary $x_+=x_K=4.197$, where $K_E=0$ 
(shaded area corresponds to $K_E<0$).
}
\label{fig11}
\end{figure}

\begin{figure}[t]
\centering
\includegraphics[width=11cm]{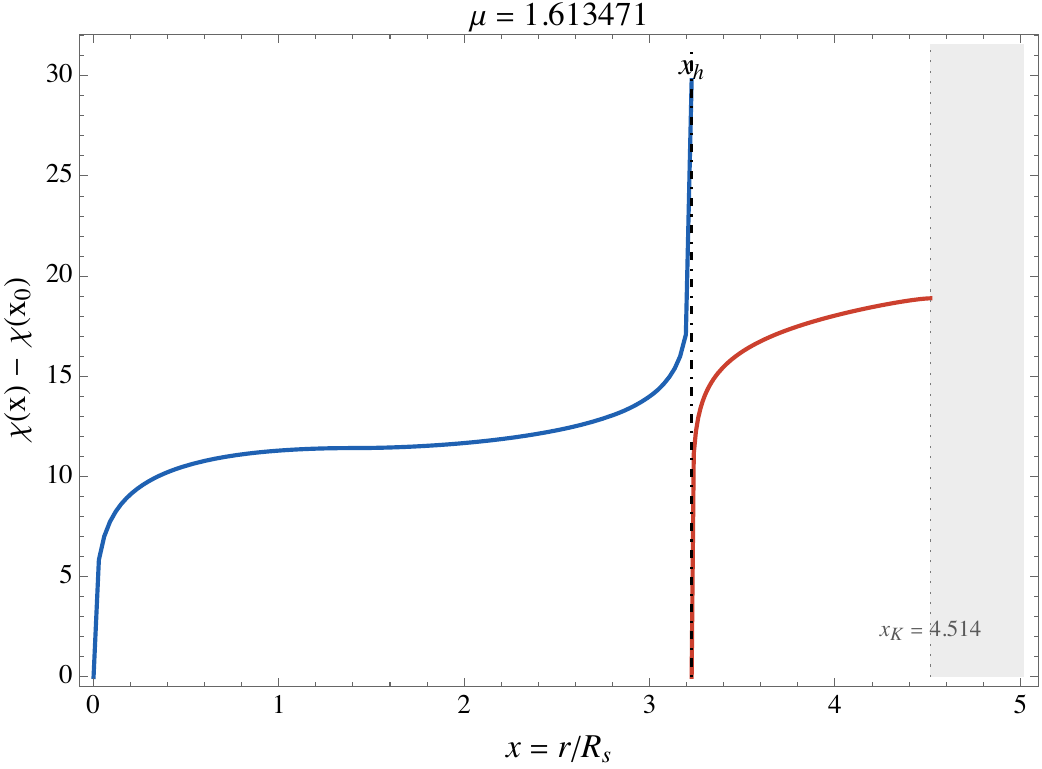}
\caption{Canonical scalar field  $\chi(x)$ for $\mu=1.613471=\mu_{\rm bif}$. 
The dash-dotted vertical line denotes a reconstruction pole, while the
dotted line marks the boundary $x_+=x_K=4.514$ where
$\dd\chi/\dd x=0$ (shaded  area corresponds to $K_E<0$).
}
\label{fig12}
\end{figure}

\begin{figure}[h]
\centering
\includegraphics[width=11cm]{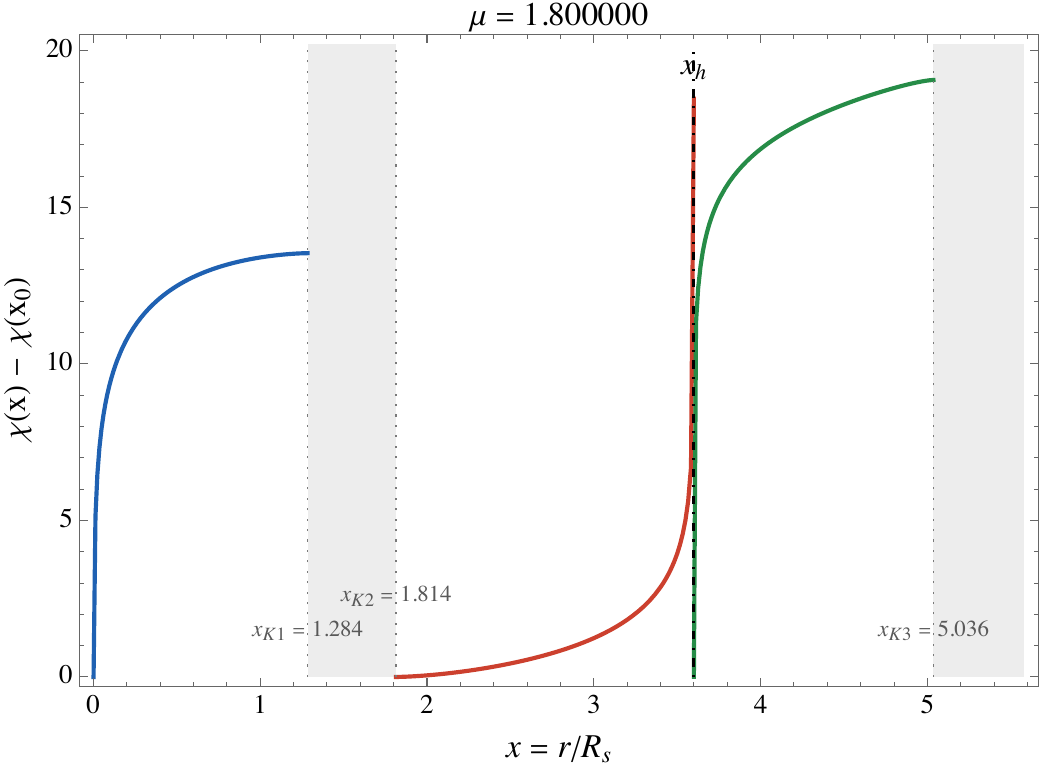}
\caption{Canonical scalar field $\chi(x)$ for $\mu=1.8>\mu_{\rm bif}$.
The kinetic boundaries $x_{K1}$ and $x_{K2}$ delimit the inner shaded region
where $K_E<0$, while $x_{K3}$ marks the outermost kinetic boundary.
The reconstruction pole at $x_Y=x_h$ separates the intermediate and outer
canonical branches.}
\label{fig13}
\end{figure}

Figs.~\ref{fig11}--\ref{fig13} show the canonical field $\chi$ on the connected
positive-kinetic branches for three representative values of $\mu$
chosen below, at, and above the bifurcation value $\mu_{\rm bif}$ in Eq.~\eqref{bif1}.
The constant $\chi_{0a}$ is fixed independently on each connected component 
$x_{a-}<x<x_{a+}$ by setting $\chi_a(x_{0a})=0$ with $x_{0a}=x_{a-}$. 

Fig.~\ref{fig11} displays the canonical field for $\mu=1.5<\mu_{\rm bif}$.
The corresponding domain $\mathcal I(\mu=1.5)$ consists of two positive-kinetic
branches, $0<x<x_Y$ and $x_Y<x<x_K$, split by a reconstruction pole at
$x_Y=3.0$ where $|Y|$ diverges.
This reconstruction pole coincides exactly with the location of the
event horizon, $x_Y=x_h(\mu)$.
The outer branch terminates at the kinetic boundary $x_K=4.197$,
where $K_E(x_K,1.5)=0$. 
As the reconstruction pole is approached, the steep growth of
$\chi(x)$ reflects the corresponding growth of the radial canonical
derivative $\dd\chi/\dd x=\phi_{,x}\sqrt{K_E}
=\sqrt{\mathcal P/\mathfrak D}$,
whereas the vanishing slope at the kinetic boundary follows directly
from $\dd\chi/\dd x\to 0$.

At the bifurcation point $\mu=\mu_{\rm bif}$, Fig.~\ref{fig12} shows an analogous 
structure, with the reconstruction pole shifted to $x_Y=x_h=3.227$ and the outer
kinetic boundary to $x_K=4.514$. 

For $\mu=1.8>\mu_{\rm bif}$, an additional qualitative feature appears in Fig.~\ref{fig13}.
There are now two regular kinetic boundaries located at $x_{K1}=1.284$ and
$x_{K2}=1.814$ which delimit the excluded interval $x_{K1}<x<x_{K2}$
where $K_E(x,1.8)<0$, shown by the lightly shaded region in Fig.~\ref{fig13}.
The two positive-kinetic components on either side consequently define
distinct real canonical branches.
A further reconstruction pole near $x_Y=x_h=3.6$ separates the intermediate
branch $x_{K2}<x<x_Y$ from the outer branch $x_Y<x<x_{K3}$ (see also
Appendix~\ref{app2} for more details).
Thus, above the kinetic bifurcation, the domain $\mathcal I$ contains
three connected regions, one near the centre, one inside the horizon
and one outside. 

\begin{figure}[h]
\centering
\includegraphics[width=11cm]{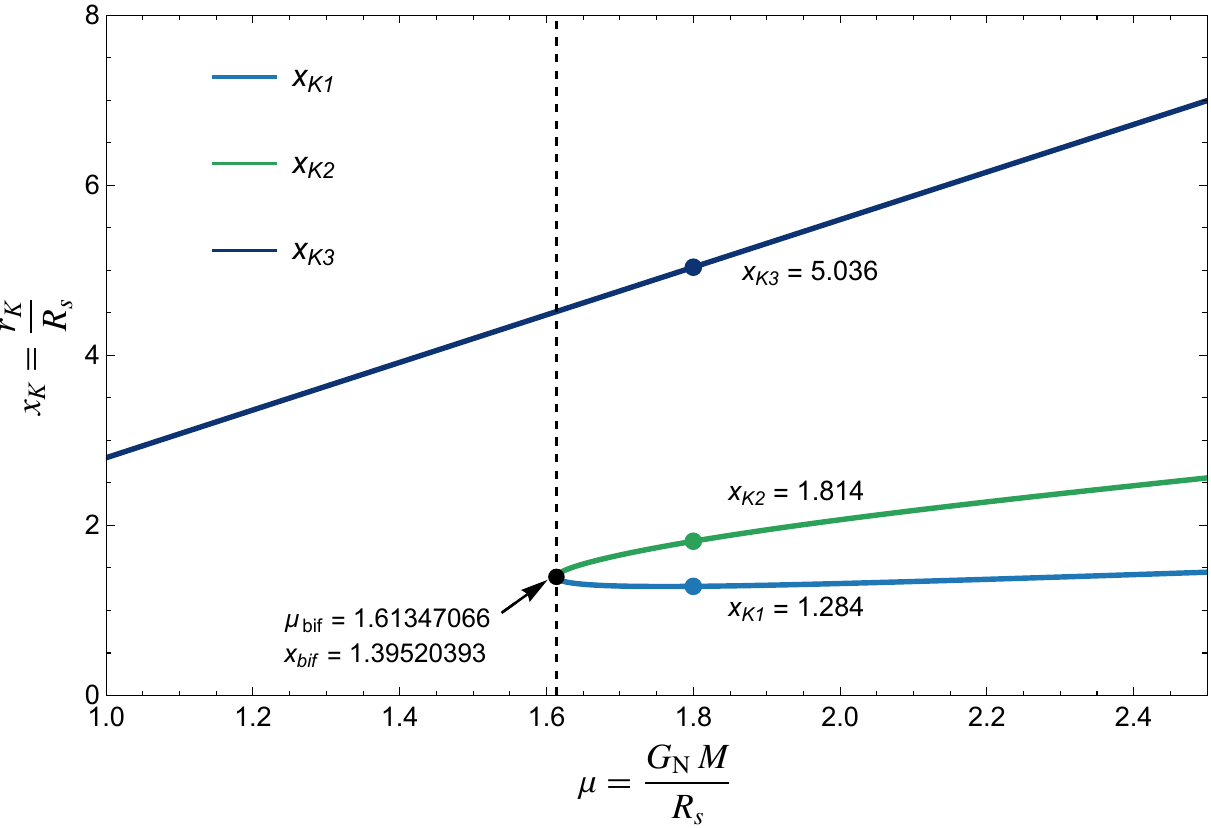}
\caption{Boundaries of the kinetic branches where $K_E(x_{Ki},\mu)=0$.
The dashed black line indicates $\mu=\mu_{\rm bif}$,
where the two inner branches coalesce at the non-degenerate fold,
$x_{K1}=x_{K2}=x_{\rm bif}$.
The dashed gray line indicates the case $\mu=1.8$ displayed in Fig.~\ref{fig13}.
}
\label{fig14}
\end{figure}

Fig.~\ref {fig14} shows the regular zeros $x_{Ki}$ of the Einstein-frame
kinetic coefficient $K_E$ as functions of $\mu$.
The critical value $\mu=\mu_{\rm bif}$ marks a fold bifurcation at $x=x_{\rm bif}$ 
where the two inner zeros $x_{K1}$ and $x_{K2}$ coalesce.
Thus, the value $\mu_{\rm bif}=1.61347066$ defines the fold at $x_{\rm bif}=1.39520393$,
where the two inner branches coalesce and, for $\mu>\mu_{\rm bif}$,
split into two distinct simple zeros, while the outer branch $x_{K3}$ remains distinct.

We remark again that the existence of lower and upper bounds $x_\pm$ of connected
regions is qualitatively in agreement with the necessary regularisation scales $R_s$
and $R_\infty$ in Eq.~\eqref{NM-regulated}.
However, it must be emphasised that the event horizon $x=x_h$
(which exists only for $\mu>\mu_h$ in Eq.~\eqref{muh}) is strictly excluded
from all the connected regions $\mathcal I_a$ since it is always a reconstruction pole,
$x_h=x_Y$, that is a location where the mapping~\eqref{Einsteintransform}
between the Jordan frame and the Einstein frame is singular.
Moreover, the connected domains which cover the region outside the horizon
never reach very large radial extensions (compared to the horizon radius),
as can be clearly appreciated from Figs.~\ref{plot1} and~\ref{plot12}.

\section{Derivative-coupled Horndeski reconstruction}
\label{sec4}

We consider the derivative-coupled scalar-tensor action
\begin{equation}
S
=
\frac{1}{16\pi \gn}\int d^4x\,\sqrt{-g}\,
\left[
\mathcal R
-Z(\phi)\nabla_\mu\phi\nabla^\mu\phi
-2U(\phi)
+H(\phi)G^{\mu\nu}\nabla_\mu\phi\nabla_\nu\phi
\right]
\ ,
\label{Haction}
\end{equation}
where the Einstein--Hilbert coefficient is fixed to unity and the last term is
the Einstein-tensor derivative coupling belonging to the Horndeski class.
We will reconstruct the three functions $Z(\phi)$, $U(\phi)$, and $H(\phi)$ on the
background metric~\eqref{metric} with the scalar field~\eqref{phi1}.
It is convenient to introduce the radial combinations
\begin{equation}
W(r)
\equiv
H[\phi(r)]
\left[\phi'(r)\right]^2
\ ,
\qquad
K(r)
\equiv
Z[\phi(r)]
\left[\phi'(r)\right]^2
\ .
\label{WKdef}
\end{equation}
Since $H(\phi)$ and $Z(\phi)$ enter the static background equations
through these combinations, the reconstruction can be carried out
directly in terms of $W(r)$, $K(r)$, and $U(r)$, without first
inverting the scalar profile.
The covariant functions $H(\phi)$ and $Z(\phi)$ are recovered subsequently
whenever $\phi(r)$ is locally monotonic.

The corresponding field equations for the metric are given by
\begin{equation}
\mathcal E_{\mu\nu}
\equiv
\frac{16\pi G}{\sqrt{-g}}
\frac{\delta S}{\delta g^{\mu\nu}}=G_{\mu\nu}
-Z(\phi)
\left(
\nabla_\mu\phi\nabla_\nu\phi
-\frac{1}{2}g_{\mu\nu}(\nabla\phi)^2
\right)
+U(\phi)g_{\mu\nu}
+\mathcal E_{\mu\nu}^{(H)}
=
0
\ ,
\label{metricELH}
\end{equation}
with the complete expression for $\mathcal E_{\mu\nu}^{(H)}$ provided in Appendix~\ref{apph1}.
The nontrivial components of Eq.~\eqref{metricELH} are
$\mathcal E^t_{\ t}=\mathcal E^r_{\ r}=\mathcal E^\theta_{\ \theta}=\mathcal E^\varphi_{\ \varphi}=0$.
These equations are not all independent, due to diffeomorphism invariance and
the scalar equation of motion, and two independent (combinations) will suffice for the background
reconstruction.

A particularly useful combination is given by
\begin{equation}
\mathcal E^t_{\ t}-\mathcal E^r_{\ r}=0
\ ,
\label{metricdiffH}
\end{equation}
in which the potential term $U$, being proportional to $\delta^\mu_{\ \nu}$, cancels identically.
Evaluating Eq.~\eqref{metricdiffH} for a metric of the form~\eqref{line-element}, with $K$ and $W$ defined
in Eqs.~\eqref{WKdef}, yields
\begin{equation}
K(r)=
\frac{2f}{r^2(f+1)}
\left[
(f-1)W-rfW'
\right]
\ ,
\label{metric-difference-H}
\end{equation}
which fixes $K$ once $f$ and $W$ are specified.
The terms proportional to $W$ and $W'$ originate from the Einstein-tensor derivative
coupling and provide the additional structure absent in the minimally coupled
two-derivative sector.

The potential $U$ can then be determined from either of the remaining independent metric
field equations.
Substituting Eq.~\eqref{metric-difference-H} into $\mathcal E^t_{\ t}=0$ gives
\begin{equation}
U(r)=
\frac{2
-3Wrf^2f'
-3Wrff'
-3Wf^3+Wf
-2rf^2W'
-2rff'
-2rf'
-2f^2}
{2r^2(f+1)}
\ .
\label{UH}
\end{equation}
With $K$ and $U$ determined by Eqs.~\eqref{metric-difference-H} and~\eqref{UH},
respectively, the remaining reconstruction condition is provided by the scalar field
equation of motion, which results in a second-order linear differential equation
for $W(r)$ which we will study below [see Eq.~\eqref{WODE}].

It is now easy to show that the coherent state geometry~\eqref{metric} cannot be recovered
in the restricted sector $H=0$. 
The latter implies $W=0$ and Eq.~\eqref{metric-difference-H} immediately gives
\begin{equation}
K(r)
=
Z[\phi(r)]
\left[\phi'(r)\right]^2
=
0
\ .
\label{minimal-obstruction-H}
\end{equation}
From Eq.~\eqref{phi1} one has $\phi'(r)\neq0$, so that $Z[\phi(r)]=0$
and the scalar field equation implies $U_{,\phi}=0$, that is $U[\phi(r)]$ must be constant.
Eq.~\eqref{UH} however requires
\begin{equation}
U(r)
=
-\frac{rf'(r)+f(r)-1}{r^2}
=
\frac{4\gn M}{\sqrt{\pi}R_s r^2}
\exp\left(-\frac{r^2}{R_s^2}\right)
\ ,
\label{minimal-U-H}
\end{equation}
which is manifestly nonconstant.
Thus, in agreement with footnote~\ref{foot2}, the coherent-state geometry cannot be
reconstructed in the restricted sector $F=1$ with $H=0$ with the assumed scalar
profile~\eqref{phi1}.
It is still important that this obstruction does not originate solely from fixing the
Einstein--Hilbert coefficient $F=1$ in Eq.~\eqref{Haction}.
In fact, for $H(\phi)\neq0$, the additional function $W(r)=H[\phi(r)]\phi'(r)^2$ lifts
this restriction and permits the reconstruction.

The second-order linear equation for $W$ reads
\begin{equation}
\mathcal A_2 W''+\mathcal A_1 W'
+\mathcal A_0 W+\mathcal S=0
\ ,
\label{WODE}
\end{equation}
where
\begin{subequations}
\begin{eqnarray}
\mathcal A_2&=&
-2r^2(f-1)(f+1)f^2
\ ,
\\
\mathcal A_1&=&
-2rf\left[
2rf^2f'+rff'-3rf'
+f^3+3f^2+f-1
\right]
\ ,
\\
\mathcal A_0
&=&
r^2f^3f''
+r^2f^2f'^2
+2r^2f^2f''
+2r^2ff'^2
+r^2ff''
+r^2f'^2
\nonumber
\\
&&
+
3rf^3f'
-3rf^2f'
-11rff'
+3rf'
-2f^4-6f^3-6f^2-2f
\ ,
\\
\mathcal S&=&
2(f+1)^2\left(r^2f''-2f+2\right)
\ .
\label{coeffH}
\end{eqnarray}
\end{subequations}
Thus the Horndeski reconstruction is completely reduced to a linear
second-order boundary-value problem.

The source term~\eqref{coeffH} can also be written as
\begin{equation}
\mathcal S(r)
=
-2r^2(f+1)^2
\left(
\mathcal R+4G^t_{\ t}
\right)
\ ,
\label{sourcecurvatureH}
\end{equation}
which makes it apparent that it is a purely geometrical forcing term generated by
the prescribed coherent-state metric.
In particular, for the classical Schwarzschild geometry~\eqref{V_N},
one has $r^2f''-2f+2=0$, hence $\mathcal S=0$ as expected.
On the other hand, for the coherent-state geometry~\eqref{metric}
in the dimensionless variables~\eqref{d1}, the source term explicitly reads
\begin{equation}
\mathcal S(x,\mu)
=
\frac{64\mu}{\sqrt{\pi}}\,
\frac{x^2+1}{x^2}
\left[
x-\mu\operatorname{erf}(x)
\right]^2
e^{-x^2}
\ ,
\label{sourceH}
\end{equation}
which is exponentially suppressed for $x\gg1$.
The inhomogeneous Horndeski deformation therefore remains localized
in the region where the coherent-state geometry departs appreciably
from the Schwarzschild vacuum.

Like in the previous scalar-tensor case, the background reconstruction guarantees that the
coherent-state geometry is an exact solution of the Horndeski theory~\eqref{Haction},
but does not ensure dynamical consistency.
In particular, the coupling to the Einstein tensor modifies the kinetic term of the
scalar field and the previous analysis of positivity for the non-minimally coupled
scalar-tensor reconstruction cannot be carried over unchanged.
In fact, we can define
\begin{equation}
\mathcal L_{\rm der}
=
-\left[
Z(\phi)g^{\mu\nu}-H(\phi)G^{\mu\nu}
\right]
\nabla_\mu\phi\nabla_\nu\phi
\equiv
-\mathcal K^{\mu\nu}\nabla_\mu\phi\nabla_\nu\phi
\ ,
\label{KtensorH}
\end{equation}
which shows that the effective kinetic operator is anisotropic whenever the background
Einstein tensor $G_{\mu\nu}$ does not vanish.

In a local orthonormal tetrad $e_a^\mu\,g_{\mu\nu}\,e^\nu_b=\eta_{ab}$
adapted to a metric $g_{\mu\nu}$ of the form given in Eq.~\eqref{line-element},
one has
\begin{equation}
\mathcal K^{\mu\nu}\,\nabla_\mu\phi\nabla_\nu\phi
=
\mathcal K^a_{\ b}\,\nabla^a\phi\nabla_b\phi
\ ,
\end{equation}
where
$\mathcal K^a_{\ b}=\operatorname{diag}\left(\mathcal K_t,\mathcal K_r,\mathcal K_\Omega,\mathcal K_\Omega\right)$,
with
\begin{subequations}
\begin{align}
\mathcal K_t(r)
&
\equiv Z-HG^t_{\ t}
=
Z-HG^r_{\ r}
\equiv 
\mathcal K_r(r)
\ ,
\\
\mathcal K_\Omega(r)
&\equiv Z-HG^\theta_{\ \theta}
\ .
\end{align}
\end{subequations}
Consequently, the derivative part of the scalar Lagrangian takes the form
\begin{equation}
\mathcal L_{\rm der}
=
\mathcal K_t(\partial_{ t}\phi)^2
-\mathcal K_r(\partial_{ r}\phi)^2
-\mathcal K_\Omega
\left[
(\partial_{\theta}\phi)^2
+(\partial_{\varphi}\phi)^2
\right]
\ .
\label{LkinorthH}
\end{equation}
The corresponding contribution to the local Hamiltonian density reads
\begin{equation}
\mathcal H_{\rm kin}
=
\mathcal K_t(\partial_{ t}\phi)^2
+\mathcal K_r(\partial_{ r}\phi)^2
+\mathcal K_\Omega
\left[
(\partial_{\theta}\phi)^2
+(\partial_{\varphi}\phi)^2
\right]
\ ,
\label{HkinorthH}
\end{equation}
from which we obtain the positivity conditions $K^a_{\ b}>0$, or
$\mathcal K_t=\mathcal K_r>0$, and $\mathcal K_\Omega>0$.

It is further convenient to express these conditions directly in terms of the radial variables
$K$ and $W$ introduced in Eq.~\eqref{WKdef}.
We therefore define the rescaled kinetic coefficients
\begin{equation}
\mathscr K_t(r)
\equiv
\left[\phi'(r)\right]^2
\mathcal K_t(r)
=
K(r)-W(r)G^t_{\ t}
=
K(r)-\frac{W(r)}{r^2}
\left[rf'(r)+f(r)-1\right]
\quad
\label{KtH}
\end{equation}
and
\begin{equation}
\mathscr K_\Omega(r)
\equiv
\left[\phi'(r)\right]^2
\mathcal K_\Omega(r)
=
K(r)-W(r)G^\theta_{\ \theta}
=
K(r)-W(r)
\left[
\frac12 f''(r)+\frac{f'(r)}{r}
\right]
\ .
\label{KOmegaH}
\end{equation}
We notice that the factor $(\phi')^2$ is strictly positive (where $\phi'(r)\neq0$)
and the rescaling does not alter the signs of the kinetic coefficients.

We also introduce the dimensionless kinetic combination
\begin{equation}
\begin{aligned}
k(x)
&
\equiv R_s^2\,K(r)
=
Z[\phi(x)]\,\phi_{,x}^2
\\
&=
\frac{
\left[x-2\mu\operatorname{erf}(x)\right]
\left\{
x\left[x-2\mu\operatorname{erf}(x)\right]
W_{,x}
+2\mu\operatorname{erf}(x)\,W
\right\}
}{
x^3\left[\mu\operatorname{erf}(x)-x\right]
}
\end{aligned}
\label{kdimH}
\end{equation}
and define the dimensionless potential
\begin{equation}
u(x)\equiv R_s^2\,U(r)
\ ,
\label{udefH}
\end{equation}
which, using Eq.~\eqref{UH}, reads 
\begin{equation}
u(x)
=
-\frac{(f+1)
\left[
3xWf f_{,x}
+2x f_{,x}
+2(f-1)
\right]
+Wf(3f^2+1)
+2xf^2W_{,x}
}{2x^2(f+1)}
\ .
\label{udimH}
\end{equation}

The positivity conditions can now be explicitly expressed in terms of the couplings
in the action~\eqref{Haction}.
For the scalar field~\eqref{phi1}, we have 
\begin{subequations}
\begin{align}
H[\phi(x)]
&
=
\frac{W(x)}{\phi'(r)^2}
=
\frac{\pi R_s^2}{4}\,
e^{2x^2}W(x)
\ ,
\label{Hrecon}
\\
Z[\phi(x)]
&
=
\frac{k(x)}{\phi_{,x}^2}
=
\frac{\pi}{4}\,
e^{2x^2}k(x)
\ ,
\label{Zrecon}
\\
U[\phi(x)]
&
=
\frac{u(x)}{R_s^2}
\ .
\label{Urecon}
\end{align}
\end{subequations}
Since the profile~\eqref{phi1} is strictly monotonic for every finite $x$, it can be inverted as
\begin{equation}
x=\operatorname{erf}^{-1}(\phi)
\ ,
\label{phiinverseH}
\end{equation}
so that Eqs.~\eqref{Hrecon}-\eqref{phiinverseH} define the functions $H(\phi)$, $Z(\phi)$,
and $U(\phi)$ explicitly on each regular reconstructed branch.
The radial quantities $W(x)$, $k(x)$, and $u(x)$ are therefore the on-shell representatives
of the three coupling functions entering the action~\eqref{Haction}.

\begin{figure}[t]
\centering
\includegraphics[width=11cm,height=9.5cm]{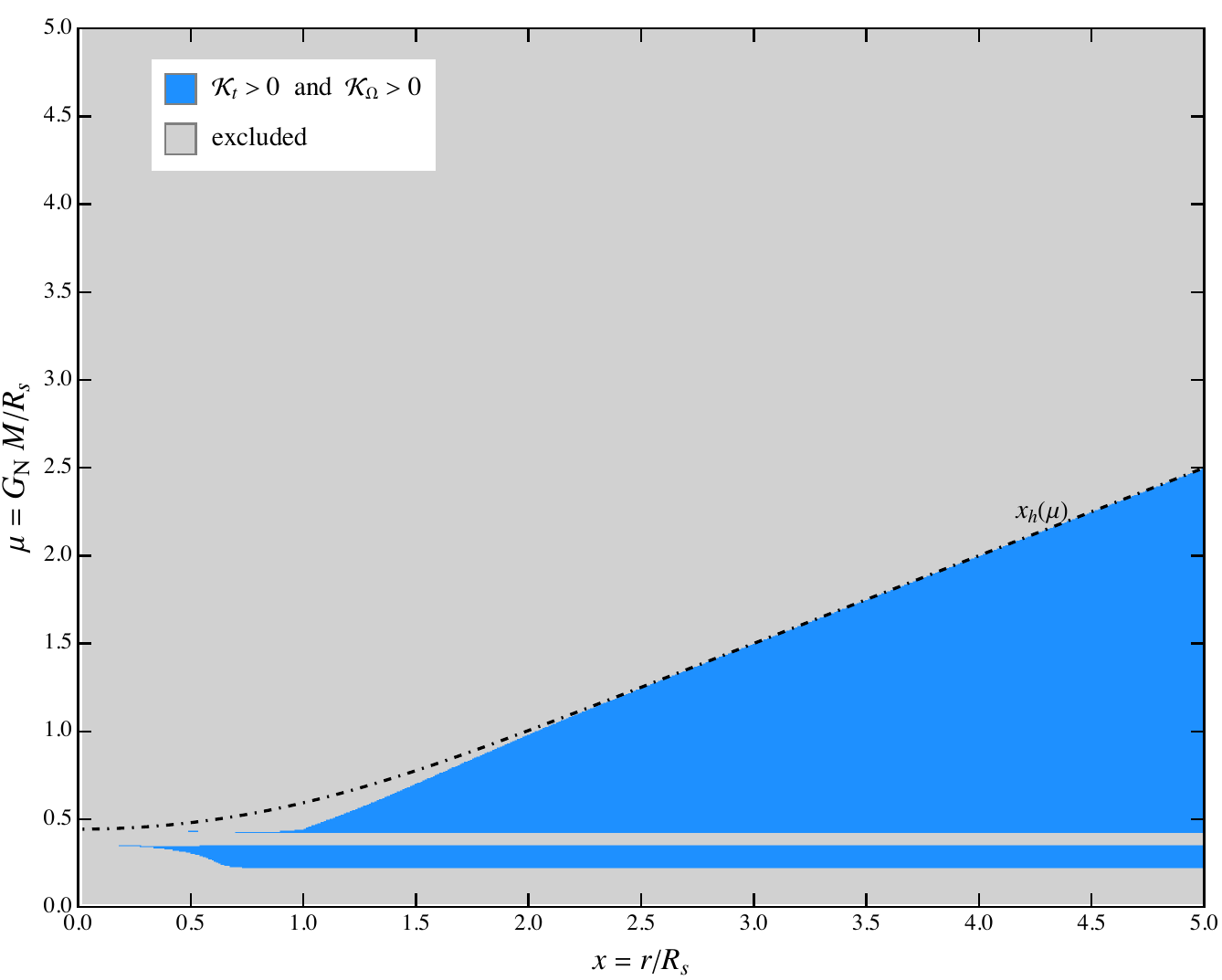}
\caption{Domain of viability for the reconstructed Horndeski theory.
The blue region satisfies simultaneously $\mathscr K_t>0$ and $\mathscr K_\Omega>0$,
whereas the gray region is excluded by at least one of these conditions.
The dash-dotted line represents the location of the event horizon.}
\label{fig17}
\end{figure}

In terms of dimensionless quantities, Eqs.~\eqref{KtH} and~\eqref{KOmegaH} respectively
become
\begin{eqnarray}
\mathscr K_t(x,\mu)
&=&
k(x)
-\frac{W(x)}{x^2}
\left(
xf_{,x}+f-1
\right)
\ ,
\label{KtdimH}\\
\mathscr K_\Omega(x,\mu)
&=&
k(x)
-W(x)
\left(
\frac{1}{2}f_{,xx}
+\frac{f_{,x}}{x}
\right)
\ .
\label{KOmegadimH}
\end{eqnarray}
These two quantities determine the sign of the scalar kinetic term and 
the necessary conditions of admissibility for the coherent state background are
therefore given by
\begin{equation}
\mathscr K_t(x,\mu)>0
\quad
{\rm and}
\quad
\mathscr K_\Omega(x,\mu)>0
\ .
\label{fixedadmissibleH}
\end{equation}
The corresponding domain, $\mathcal I_H=\left\{(x,\mu)\,|\,\mathscr K_t>0,\ \mathscr K_\Omega>0\right\}$,
is illustrated in Fig.~\ref{fig17}.

An important feature shown in Fig.~\ref{fig17} is that the action~\eqref{Haction} can be used
to recover the coherent state background in large portions of the spacetime outside the event
horizon but not its interior.
This is quite the opposite of what we found for the scalar-tensor theory~\eqref{Einsteinaction}. 
Moreover, it can be established analytically that the horizon $x=x_h$
determined by $f(x_h,\mu)=0$ in Eq.~\eqref{mx} (or, equivalently, Eq.~\eqref{Dzero1}) is always
excluded.
Since the denominator in Eq.~\eqref{kdimH} equals $x_h-\mu\operatorname{erf}(x_h)=x_h/2\neq0$,
one finds $k(x_h)=0$ provided $W$ and $W_{,x}$ do not vanish at $x=x_h$.
Direct differentiation of Eq.~\eqref{mx} yields
\begin{equation}
xf_{,x}+f-1
=
-\frac{4\mu}{\sqrt{\pi}}e^{-x^2}
=
-\frac{f_{,xx}}{2}-\frac{f_{,x}}{x}
\ ,
\label{horizoncurvatureH}
\end{equation}
which, on substituting into Eqs.~\eqref{KtdimH} and~\eqref{KOmegadimH}, yields
\begin{equation}
\mathscr K_t(x_h,\mu)
=
\frac{4\mu e^{-x_h^2}}{\sqrt{\pi}x_h^2}W(x_h)
=
-\frac{\mathscr K_\Omega(x_h,\mu)}{x_h^2}
\ ,
\label{horizoncoefficientsH}
\end{equation}
so that
\begin{equation}
\mathscr K_t(x_h,\mu)
\mathscr K_\Omega(x_h,\mu)
=
-\frac{16\mu^2e^{-2x_h^2}}{\pi x_h^2}
W(x_h)^2
\leq
0
\ .
\label{horizonproductH}
\end{equation}
This result is incompatible with the simultaneous strict positivity~\eqref{fixedadmissibleH}.
Therefore, $x_h\notin\mathcal I_H$ for any value of the black hole mass.

\section{Conclusions}
\label{sec5}
The main result of this work is that the semiclassical black hole geometry obtained from coherent
states admits exact scalar-tensor realisations whose kinetic consistency dynamically selects
the physically admissible domains.
The necessary regularisation of the coherent state at low as well as high momenta could therefore
be related to the dynamical viability of the scalar-tensor realisation.
However, such realisation does not appear to be able to consistently reproduce the whole semiclassical
spacetime and the location of the event horizon, in particular, is always excluded.~\footnote{This
exclusion is reminiscent of the coherent state approach developed in Ref.~\cite{Bambagiottiab:2026tnr}.}

For the non-minimally coupled scalar-tensor reconstruction, positivity of the kinetic coefficient
in the Einstein frame reveals finite admissible intervals, reconstruction boundaries, and a kinetic
bifurcation controlled by the mass-to-core scale ratio.
Above the critical value, a degenerate kinetic zero splits into two simple zeros which bound
an excluded interval.
In addition, the asymptotic negativity of the kinetic coefficient at spatial infinity implies 
that the scalar-tensor realisation is intrinsically restricted to finite radial domains.
We further show that the minimally coupled sector is incompatible with the coherent-state
geometry and non-constant scalar field, thereby requiring a non-minimal coupling
in the Jordan-frame reconstruction.

The canonical Einstein-frame description provides the corresponding field-space interpretation.
Each connected positive-kinetic region defines an independent canonical scalar branch,
while kinetic boundaries mark the loss of local invertibility of the canonical field redefinition.
These kinetic boundaries are distinct from the reconstruction poles,
which arise in the Jordan-frame inverse reconstruction and independently separate
regular branches.

We have further shown that the same coherent-state geometry admits an exact realisation
within a derivative-coupled Horndeski theory with a fixed Einstein--Hilbert coefficient and
an Einstein-tensor derivative coupling.
The corresponding two-derivative minimally coupled sector cannot support the prescribed
geometry for a non-constant scalar profile, whereas the derivative coupling removes this
obstruction.
The reconstruction then reduces to a second-order linear boundary-value problem driven
by a purely geometric source associated with the departure of the coherent-state geometry
from the Schwarzschild vacuum.

The Horndeski realisation also modifies the conditions of kinetic admissibility.
The derivative coupling makes the effective scalar kinetic operator anisotropic.
The simultaneous positivity of temporal, radial, and angular derivative terms determines
the background kinetically admissible regions.
Their boundaries are determined directly by the reconstructed Horndeski action and
are distinct from those governing the non-minimal scalar-tensor realisation.

The two reconstructions show that the dynamical-cutoff mechanism is not constrained
to a particular scalar-tensor parametrisation.
In the non-minimal theory, it is encoded by the sign and bifurcation structure of the
Einstein-frame kinetic sector, whereas in the derivative-coupled Horndeski theory,
it arises from the anisotropic effective kinetic operator.
In both cases, the admissible domains are selected dynamically
after reconstruction of the target geometry.

The fixed-background Horndeski conditions should nevertheless be regarded
as necessary rather than sufficient stability criteria.
A complete analysis requires the coupled scalar-metric perturbation system
and the associated no-ghost and gradient-stability conditions.
This provides a natural direction for further investigation, together with the global
branch structure of the Horndeski reconstruction and the dependence of its kinetic
boundaries on the characteristic scales of the geometry.
More generally, our results show that inverse gravitational reconstruction
provides not only a systematic route to covariant theories supporting prescribed
quantum-corrected geometries, but also dynamically identify the spacetime domains
over which their effective scalar-tensor descriptions remain physically admissible.

\subsection*{{Acknowledgments}} 
R.C.~and A.G.~are partially supported by the INFN grant FLAG and their work has also been carried out
in the framework of activities of the National Group of Mathematical Physics (GNFM, INdAM).
A.G.~is partially supported by the Italian Ministry of Universities and Research (MUR) through the grant
``BACHQ: Black Holes and The Quantum'' (grant no.~J33C24003220006).
R.d.R.~thanks DIFA-UniBo for the hospitality, the S\~ao Paulo Research Foundation -- FAPESP
(Grants No.~2021/01089-1, No.~2025/23004-9, and No.~2026/14943-4),
and to the National Council for Scientific and Technological Development -- CNPq
(Grants No.~303742/2023-2 and No.~401567/2023-0), for partial financial support.

\appendix

\section{Asymptotic origin of the large kinetic boundary}
\label{app2}

We provide here additional details on the largest kinetic boundary $x_{K}\gg 1$ at which
$K_E(x,\mu)$ changes sign.
On every regular reconstruction branch, $\mathfrak D(x,\mu)>0$ away from its zeros.
Moreover, the reparametrisation factor $(\pi/4)e^{2x^2}$ in Eq.~\eqref{KEdimensionless} is strictly
positive for every finite $x$, hence $\operatorname{sgn}K_E(x,\mu)=\operatorname{sgn}\mathcal P(x,\mu)$
at every regular point.
Also, for fixed $\mu>0$, the asymptotic expansion in Eq.~\eqref{KElargex}
shows that there exists a finite value $X(\mu)$ such that $K_E(x,\mu)<0$ for all $x>X(\mu)$.
Therefore, $K_E(x_0,\mu)>0$ at some finite $x_0<X(\mu)$, together with the asymptotic negativity of
$K_E(x,\mu)$ for $x\gg 1$, will analytically establish the existence of a kinetic boundary $x_{K}(\mu)$
for all $\mu>0$, provided both points belong to the same connected regular $\mathcal I_a$.

For $\mu>\sqrt{\pi}/2$, the existence of such an $x_0$ follows directly from Eq.~\eqref{KEsmallx},
since $K_E>0$ in a sufficiently small (punctured) neighbourhood of the origin that
belongs to a regular branch.
On the other hand, for $0<\mu\ll 1$, the numerator~\eqref{PKE} can be expanded as
\begin{equation}
\mathcal P(x,\mu)
=
4\sqrt{\pi}\,\mu e^{-x^2}x^6
(x^2-1)(2x^2+1)
+\mathcal O(\mu^2)
\ .
\end{equation}
Consequently, for any fixed $x_0>1$, $\mathcal P(x_0,\mu)>0$ for sufficiently small $\mu>0$.
Since $\mathfrak D>0$ at regular points, $K_E(x_0,\mu)>0$.
Combined with the asymptotic negativity of $K_E$, continuity then implies the existence 
of a finite kinetic boundary $x_{K}$ on the corresponding connected regular branch,
unless a reconstruction pole is encountered first.
For any remaining compact interval in $\mu$, the existence and regularity of $x_K$
can be verified directly from the exact expressions~\eqref{PKE} and~\eqref{dx1}
by identifying a positive value of $\mathcal P$ followed, on the same regular branch,
by a negative value.
The intermediate value theorem then guarantees the existence of a finite regular zero of $K_E$.
Reconstruction poles remain independent boundaries and must
be excluded when identifying such a zero with $x_{K}$.

The scaling of the outer kinetic boundary can also be obtained analytically.
For large $\mu$, we consider the double-scaling limit $\mu\to\infty$ with $c\equiv x/\mu$ fixed. 
In the regime where the exponentially suppressed terms are negligible, one has
\begin{equation}
\mathcal P(x,\mu)
=
\pi\mu^6\,Q(x/\mu)
+\mathcal O\!\left(e^{-c^2\mu^2}\mu^5\right)
\ ,
\end{equation}
with
\begin{equation}\label{qc}
Q(c)
=
32-96c+66c^2+4c^3-6c^4
=
-2(c-1)(3c^3+c^2-32c+16)
\ .
\end{equation}
The relevant root of $Q(c)=0$ for $2<c<3$ reads $c_\star\simeq 2.7978232765$. 
It follows that the outer kinetic boundary asymptotically satisfies
\begin{equation}
x_{K3}(\mu)
\simeq
c_\star\mu
\simeq
2.7978232765\,\mu
\ ,
\end{equation}
in agreement with the numerical evolution found in Sec.~\ref{subsec3}.

\section{Regular bifurcation point and nondegenerate fold}
\label{app1}

We can establish that the numerical candidate $(x_{\rm bif},\mu_{\rm bif})$ in
Eq.~\eqref{bif1} indeed corresponds to a regular bifurcation point
and determine the local character of the bifurcation by evaluating
the kinetic coefficient, its relevant derivatives, and the reconstruction
denominators at $(x_{\rm bif},\mu_{\rm bif})$.
Direct substitution into Eq.~\eqref{KEdimensionless} gives
$K_E(x_{\rm bif},\mu_{\rm bif})\simeq K_{E,x}(x_{\rm bif},\mu_{\rm bif})\simeq 0$,
whereas
\begin{equation}
K_{E,xx}(x_{\rm bif},\mu_{\rm bif})
\simeq 161.174
\ ,
\qquad
K_{E,\mu}(x_{\rm bif},\mu_{\rm bif})
\simeq -31.3745
\ .
\label{bifderivatives}
\end{equation}
The two factors in Eq.~\eqref{KEdimensionless} at the same point evaluate to
\begin{equation}
x_{\rm bif}
-2\mu_{\rm bif}\operatorname{erf}(x_{\rm bif})
\simeq
-1.67529
\neq
0
\end{equation}
and
\begin{equation}
\sqrt{\pi}\,x_{\rm bif}
+2\mu_{\rm bif}x_{\rm bif}e^{-x_{\rm bif}^2}
-3\sqrt{\pi}\,\mu_{\rm bif}
\operatorname{erf}(x_{\rm bif})
\simeq
-5.04778517
\neq
0
\ ,
\end{equation}
which confirms that the bifurcation does not coincide with a reconstruction singularity.
Furthermore, $K_{E,xx}(x_{\rm bif},\mu_{\rm bif})\neq 0$ and $K_{E,\mu}(x_{\rm bif},\mu_{\rm bif})\neq 0$,
showing that the critical solution is a nondegenerate fold of the zero set of $K_E(x,\mu)$.

More explicitly, expanding $K_E$ locally about the critical point gives
\begin{equation}
K_E(x,\mu)
=
\frac{1}{2}
K_{E,xx}\big|_{\rm bif}
(x-x_{\rm bif})^2
+
K_{E,\mu}\big|_{\rm bif}
(\mu-\mu_{\rm bif})
+\cdots,
\label{KEfoldexpansion}
\end{equation}
since both $K_E$ and $K_{E,x}$ vanish at the bifurcation point.
Therefore, the nearby zeros satisfy
\begin{equation}
(x-x_{\rm bif})^2
\simeq
-\frac{2K_{E,\mu}|_{\rm bif}}
{K_{E,xx}|_{\rm bif}}
(\mu-\mu_{\rm bif})
\simeq
0.3893\,(\mu-\mu_{\rm bif})
\ .
\label{foldzeros}
\end{equation}
Since the coefficient on the right-hand side is positive, the degenerate zero at
$\mu=\mu_{\rm bif}$ splits into a pair of simple zeros for $\mu>\mu_{\rm bif}$.

For $\sqrt{\pi}/{2}<\mu<\mu_{\rm bif},$ the kinetic coefficient is positive
in a connected neighbourhood of the origin and remains positive up to the first boundary of the
regular reconstructed branch.
For $\mu>\mu_{\rm bif}$, the additional pair of zeros of $K_E$ generates an intermediate interval
with $K_E<0$ and splits the positive-kinetic domain into additional disconnected components.
These kinetic boundaries must be combined with the reconstruction singularities~\eqref{Dzero1}
and~\eqref{Dzero2} and with the condition $F>0$ to determine the complete domain~\eqref{s1}.

\section{Global analysis of kinetic-fold bifurcations}
\label{app3}

We address here the possibility of additional regular kinetic folds outside the finite domain
explored numerically in Sec.~\ref{subsec3}.
We first recall that the metric~\eqref{metric} admits an event horizon only for
$\mu>\mu_h\simeq 0.44311346$ given in Eq.~\eqref{muh}.
Given the range of the numerical search in Sec.~\ref{sec3}, we just need
to consider the large $\mu$ and large $x$ regimes.
Writing the numerator of Eq.~\eqref{PKE} as
\begin{equation}
\mathcal Q(x,\mu)=\sum_{j=1}^{6}a_j(x)\mu^j,
\qquad
a_6(x)=32\pi\operatorname{erf}^6(x)>0,
\end{equation}
one obtains, for $\mu\ge100$,
\begin{equation}
\frac{\mathcal Q(x,\mu)}{\mu^6}
\ge
a_6(x)
-
\sum_{j=1}^{5}
\frac{|a_j(x)|}{100^{6-j}}
\ .
\end{equation}
Using the exact coefficients of Eq.~\eqref{PKE}, we then obtain
\begin{equation}
\sup_{0<x\le27}
\left[
\frac{1}{a_6(x)}
\sum_{j=1}^{5}
\frac{|a_j(x)|}{100^{6-j}}
\right]
\simeq 0.963813<1
\ ,
\end{equation}
and therefore $\mathcal Q(x,\mu)>0,$ in the range $0<x\le27$ and $\mu\ge100$. 
Therefore, no kinetic zero, and hence no fold, can occur in this region.

For the remaining asymptotic region, the analysis of Appendix~\ref{app2}
shows that the kinetic-zero branches approach the roots of the polynomial
$Q(c)$ in Eq. (\ref{qc}), with $c=x/\mu$.
Besides the outer root $c_\star$ discussed there, the positive roots are given by 
\begin{equation}
c_1
\simeq
0.521831394
\ ,
\qquad
c_2
=
1
\ ,
\qquad
c_3
\simeq
2.797823277
\ ,
\end{equation}
for which
\begin{equation}
Q'(c_1)
\simeq
-27.2609
\ ,
\qquad
Q'(c_2)
=
24
\ ,
\qquad
Q'(c_3)
\simeq
-158.374
\ .
\end{equation}
Hence all asymptotic zeros are simple and cannot generate a fold at the large-$x$ 
kinetic boundary.
The Gaussian corrections are exponentially suppressed there.
In particular, for $x\ge30$, $e^{-x^2}\le e^{-900}\simeq 1.36\times10^{-391}$.
The remaining asymptotic branch, $\mu(x)\simeq \frac{4}{3\sqrt{\pi}}x^6e^{-x^2}$,
approaches $\mu=0$ and is therefore excluded by Eq.~\eqref{muh}.

The above estimates constrain the possible locations of additional kinetic folds.
In particular, the large-$\mu$ bound excludes zeros of the kinetic term in the
specified parameter region, while the simplicity of the roots of $Q(c)$ characterises
the asymptotic kinetic-zero branches. 

\section{Derivative Horndeski contribution to the metric equations}
\label{apph1}

For completeness, we show here the explicit contribution of the derivative Horndeski
interaction to the metric field equations used in Sec.~\ref{sec4}.
The corresponding term in the action \eqref{Haction} reads 
\begin{equation}
S_H
=
\frac{1}{16\pi \gn}
\int d^4x\,\sqrt{-g}\,
H(\phi)G^{\mu\nu}
\nabla_\mu\phi\nabla_\nu\phi .
\label{SHapp}
\end{equation}
Defining $X_{\mu\nu}\equiv \nabla_\mu\phi\,\nabla_\nu\phi$ and
$X\equiv g^{\mu\nu}X_{\mu\nu} \equiv (\nabla\phi)^2$, the contribution of Eq.~\eqref{SHapp}
to the metric field equations, with the conventions of Eq.~\eqref{metricELH}, is given by
\begin{align}
\mathcal E_{\mu\nu}^{(H)}
=
&
-\frac{1}{2}H(\phi)
\left[
-\frac{1}{2}g_{\mu\nu}
G^{\alpha\beta}X_{\alpha\beta}
+2G_{(\mu}{}^{\alpha}X_{\nu)\alpha}
-\frac{1}{2}R X_{\mu\nu}
+R_{\mu\alpha\nu\beta}X^{\alpha\beta}
+\nabla_\mu\nabla^\alpha\phi\,
 \nabla_\nu\nabla_\alpha\phi
\right.\nonumber\\
&\left.\hspace{1.0cm}
-(\Box\phi)\nabla_\mu\nabla_\nu\phi+\frac{1}{2}g_{\mu\nu}
\left[
(\Box\phi)^2
-\nabla_\alpha\nabla_\beta\phi\,
 \nabla^\alpha\nabla^\beta\phi
\right]
\right]
\nonumber
\\
&-\frac{1}{2}H_{,\phi}(\phi)
\left[
X\,\nabla_\mu\nabla_\nu\phi
-2\nabla^\alpha\phi\,
 \nabla_\alpha\nabla_{(\mu}\phi\,
 \nabla_{\nu)}\phi
+g_{\mu\nu}
\left(
\nabla^\alpha\phi\,\nabla^\beta\phi\,
\nabla_\alpha\nabla_\beta\phi
-X\Box\phi
\right)
\right]
\ ,
\label{EHexplicit}
\end{align}
where $\Box\phi\equiv\nabla_\alpha\nabla^\alpha\phi$.

\bibliography{reconstruction.bib}
\end{document}